\documentclass[useAMS,usenatbib]{mn2e}
\usepackage{times}
\usepackage{graphics,epsfig}
\usepackage{graphicx}
\usepackage{amssymb}
\usepackage{txfonts}

\newcommand{\kms}{km\,s$^{-1}$} 
\newcommand{\Ts}   {\ensuremath{\langle {\rm T}_{\rm s} \rangle}}
\newcommand{\Tsmax}   {\ensuremath{ \langle {\rm T_s} \rangle_{max} }}

\newcommand{\Tk}   {\ensuremath{{\rm T}_{\rm k}}}
\newcommand{\Tb}   {\ensuremath{{\rm T}_{\rm B}}}

\newcommand{\cm}{cm$^{-2}$}
\newcommand{\hi}{H\,{\sc i}}
\newcommand{\hii}{H\,{\sc i}-21\,cm}

\title[Temperature of the ISM]{The temperature of the diffuse H\,{\sc i} in the Milky Way I: High resolution H\,{\sc i} 21cm absorption studies}
\author[N. Roy et al.]{Nirupam Roy$^{1}$\thanks{E-mail: nirupam@mpifr-bonn.mpg.de~(NR);~~~ nkanekar@ncra.tifr.res.in~(NK); rbraun@atnf.csiro.au (RB); chengalu@ncra.tifr.res.in~(JNC)}, 
Nissim Kanekar$^{2}$\footnotemark[1], Robert Braun$^{3}$\footnotemark[1] and Jayaram N. Chengalur $^{2}$\footnotemark[1]\\
$^{1}$Max-Planck-Institut f\"{u}r Radioastronomie, Auf dem H\"{u}gel 69, D-53121, Bonn, Germany\\
$^{2}$National Centre for Radio Astrophysics, TIFR, Post Bag 3, Ganeshkhind, Pune 411 007, India\\
$^{3}$SKA Organization, Jodrell Bank Observatory, Lower Withington, Macclesfield, Cheshire, SK11 9DL, UK}

\begin{document}
\date{Accepted yyyy month dd. Received yyyy month dd; in original form yyyy month dd}

\pagerange{\pageref{firstpage}--\pageref{lastpage}} 
\pubyear{2013}

\maketitle

\label{firstpage}

\begin{abstract}

We have carried out deep, high velocity resolution, interferometric Galactic 
\hii\ absorption spectroscopy towards 32 compact extra-galactic radio sources 
with the Giant Metrewave Radio Telescope (GMRT) and the Westerbork Synthesis 
Radio Telescope (WSRT). The optical depth spectra for most sources have root 
mean square noise values $\lesssim 10^{-3}$ per 1~\kms\ velocity channel and 
are thus sufficiently sensitive to detect absorption by warm neutral hydrogen 
with H{\sc i} column densities $N_{\rm HI} \gtrsim 10^{20}$~cm$^{-2}$, spin 
temperatures ${\rm T_s} \leq 5000$\,K, and line widths equal to the thermal 
width (20~\kms). H{\sc i}~21cm absorption was detected against all background 
sources but one, B0438$-$436. The spectra of sources observed separately with 
GMRT and WSRT show excellent agreement, indicating that spectral baseline 
problems and contamination from H{\sc i}~21cm emission are negligible. This 
paper presents the absorption spectra, the emission spectra along neighbouring 
sightlines from the Leiden-Argentine-Bonn survey, and the derived spin 
temperature spectra. On every sightline, the maximum spin temperature detected 
(at $\ge 3\sigma$ significance) even at a velocity resolution of 1~km~s$^{-1}$ 
is $\gtrsim 1000$~K, indicating that we are detecting the warm neutral medium 
along most sightlines. This is by far the largest sample of Galactic 
H{\sc i}~21cm absorption spectra of this quality, providing a sensitive probe 
of physical conditions in the neutral atomic ISM.

\end{abstract}


\section{Introduction}
\label{sec:intro}

Neutral hydrogen (\hi) in the Galactic interstellar medium (ISM) can exist 
over a range of temperatures, $\approx 20-10^4$~K. In the ``classic'' 
steady-state models of the ISM \citep[e.g.][]{field65,field69}, two distinct 
stable \hi\ phases co-exist in pressure equilibrium, a cold, high-density ($n 
\approx 10 - 100$ cm$^{-3}$) phase, the cold neutral medium (CNM) and a warm 
diffuse, low-density ($n \approx 0.1 - 1$ cm$^{-3}$) phase, the warm neutral 
medium (WNM). While the exact range of temperatures in the two temperature 
phases depends on the details of heating and cooling processes as well as the 
extent of self-shielding, typical models yield stable CNM temperatures of 
$\approx 20-400$\,K and stable WNM temperatures of $\approx 4000-9000$\,K 
\citep[e.g.][]{wolfire95,wolfire03}. In such two-phase models, gas at 
intermediate temperatures is expected to be unstable and to migrate to either 
the CNM or the WNM phases. 

\begin{figure}
\begin{center}
\includegraphics[scale=0.33, angle=-90.0]{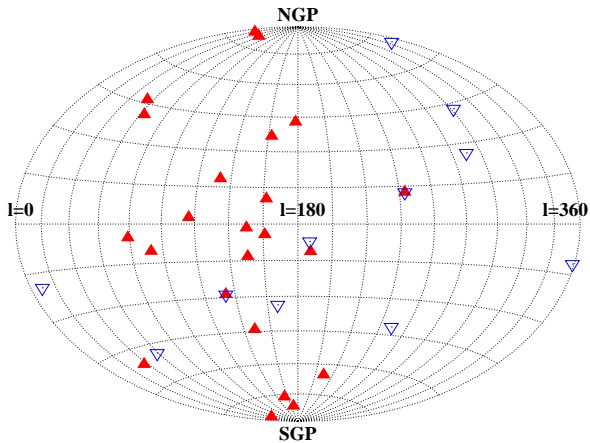}
\caption{\label{fig:obs} Observed lines of sight for the \hii\ absorption-line 
survey shown in the Galactic coordinate system. The filled and empty triangles 
are for the sightlines observed with the GMRT and the WSRT, respectively.}
\end{center}
\end{figure}

Spectroscopy in the \hii\ transition has long provided our best window on 
conditions in the neutral ISM \citep[e.g.][]{kulkarni88,dickey90}. It is well 
known that \hii\ {\it emission} studies are sensitive to the presence of both 
cold and warm gas: in the optically-thin case, the observed \hii\ brightness 
temperature is simply proportional to the total \hi\ column density. 
Conversely, \hii\ {\it absorption} studies are biased towards cold gas, as the 
\hii\ optical depth is inversely proportional to the gas temperature: for a 
given \hi\ column density, it is hence far easier to detect the CNM in 
absorption than the WNM. Such absorption studies allow one to infer the CNM 
temperature (or, given the possibility of non-thermal broadening, upper limits 
to the CNM temperature) from the widths of individual absorption components 
along a sightline. Over the last few decades, a number of \hii\ studies have 
established that the ISM indeed contains gas in the expected CNM temperature 
range $\approx 20 - 300$\,K, consistent with theoretical expectations 
\citep[e.g.][]{clark62,rad72,dickey78,heiles03a,roy06}. The high number 
densities in the CNM also indicate that the \hii\ transition is likely to be 
thermalized in this phase, with the \hii\ excitation temperature (the ``spin 
temperature'') approximately equal to the kinetic temperature 
\citep[e.g.][]{kulkarni88}. For the WNM, the relation between the spin and 
kinetic temperatures is more complicated; however, in general, the spin 
temperature is expected to be lower than the kinetic temperature here 
\citep[e.g.][]{liszt01}.

Unfortunately, the difficulty of detecting the WNM in \hii\ absorption has 
meant that relatively little is observationally known about physical 
conditions in the WNM. In recent years, two different approaches have been 
taken to address this issue. \citet{heiles03a} carried out single-dish \hii\ 
emission-absorption studies of a number of sightlines with the Arecibo 
telescope and modelled the resulting \hii\ spectra as thermally-broadened 
Gaussians to infer the kinetic temperature and the gas fraction in different 
phases along each sightline. From this, they concluded that significant 
fractions of the \hi\ lie in the thermally unstable temperature range. An 
alternative approach has been to use deep \hii\ absorption spectroscopy of 
bright compact continuum sources with long baseline interferometers, to 
attempt to directly detect the WNM in absorption and then infer physical 
conditions in this phase. The \hii\ absorption spectrum can then be modelled 
to derive conditions in the WNM. \citet{carilli98b} and \citet{dwaraka02} used 
such interferometric studies to detect the WNM in absorption towards Cygnus~A 
and 3C147, by estimating the WNM spin temperature. A somewhat different 
approach was used by \citet{kanekar03b}, based on high velocity resolution 
Australia Telescope Compact Array (ATCA) interferometric studies of two 
compact sources; they used a Gaussian decomposition procedure on the \hii\ 
absorption spectrum alone and also found a significant fraction of gas to be 
in the unstable thermal phase (a similar approach was followed by 
\citet{lane00} and \citet{kanekar01b} in cosmologically-distant damped 
Lyman-$\alpha$ systems). Later, \citet{braun05} obtained deep Westerbork 
Synthesis Radio Telescope (WSRT) absorption spectra of four targets, again 
aiming to detect the WNM in absorption. These authors modelled the \hii\ 
absorption and total-power \hii\ emission spectra with a spherically-symmetric 
isobaric two-phase model, allowing the number density and temperature to vary 
with radius.

The primary advantage of the above interferometric \hii\ absorption studies is 
that they resolve out the foreground \hii\ emission and thus yield an 
uncontaminated measure of the \hii\ absorption profile that traces gas in the 
narrow beam subtended by the compact background source. In addition, it is 
possible to obtain a high spectral dynamic range with modern interferometers 
\citep[with careful calibration of the instrumental passband; 
e.g.][]{dickey83,kanekar03b,braun05}, allowing the possibility of detecting 
the WNM in absorption. We have hence begun a large project to study the 
neutral atomic gas in the ISM through interferometric \hii\ absorption 
studies. In this paper, we present the Galactic \hii\ absorption spectra 
towards 32 compact radio-loud quasars obtained from deep, high spectral 
resolution interferometric \hii\ observations with the Giant Metrewave Radio 
Telescope (GMRT) and the WSRT. We combine these spectra (and the spectra of 
\citealt{kanekar03b}) with emission spectra on neighbouring sightlines 
obtained from the Leiden-Argentine-Bonn (LAB) survey to derive spin 
temperature spectra for 34 sightlines, at a velocity resolution of $\sim 
1$~\kms. We also present estimates of various integrated quantities along each 
sightline, including the \hi\ column density, the integrated \hii\ optical 
depth and the spin temperature. The detailed modelling of the \hii\ 
absorption/emission spectra to derive physical conditions in the ISM will be 
described in later papers. 

\begin{table}
\caption{Summary of the GMRT and the WSRT observations}
\begin{center}
 \begin{tabular}{lcc}
 \hline
Telescope            &    GMRT                &    WSRT                \\
Project ID           & 11NRb01, 12NRb01       & R05B/025, R07B/011 \\ 
No. of sources       &    $11$                &    $23$                \\
Bandwidth            &    $0.5$\,MHz          &    $2.5$\,MHz           \\
Velocity coverage    & $\sim 105$\,\kms\      & $\sim 525$\,\kms\ 		\\
No. of channels      &    $256$               &   $2048^b$               \\
Velocity resolution  & $\sim 0.4$\,\kms\       & $\sim 0.26$\,\kms$^b$   \\
Time per source$^a$  &  $10$\,hr              &  $12-24$\,hr               \\
\hline
\end{tabular}\\
\end{center}
\begin{flushleft}
$^a$~Total observing time, including all overheads.\\
$^b$~For two sources (B1328+254, and B1328+307), the WSRT observations were 
carried out with 1024 channels over a 2.5~MHz bandwidth, yielding a velocity 
resolution of $\approx 0.52$~\kms.
\end{flushleft}
\label{table:obs}
\end{table}

\section{The sample, observations and data analysis}
\label{sec:sample}

\subsection{The sample}

The main aim of this project was to detect the WNM in absorption and to 
determine its range of kinetic temperatures and the temperature distribution 
of neutral gas in the ISM. The kinetic temperature of stable WNM in the Galaxy 
is expected to lie in the range $5000 - 8000$\,K \citep{wolfire95}; this 
corresponds to line FWHMs of $\lesssim 20$~\kms. Further, the spin temperature 
of the WNM is expected to lie in the range $1000 - 5000$\,K \citep{liszt01}. 
We hence planned the observations to detect, at $5\sigma$ significance, \hii\ 
absorption with a line FWHM of 20~\kms\ and a spin temperature $\leq 5000$\,K. 
For an \hi\ column density of $2\times 10^{20}$\,\cm, this implies a peak 
optical depth of $\sim 10^{-3}$ and a spectral dynamic range of $\sim 5000$ to 
1, at a velocity resolution of 20~\kms. We also aimed to have our sightlines 
sample a large fraction of the sky, so that they sample a range of physical 
conditions. We targeted compact radio quasars, so that the absorption profile 
is obtained from a narrow pencil beam towards the radio emission. Finally, the 
use of phase calibrators as target sources simplified the observations, 
ensuring that maximum time could be spent on the targets.

Based on the above criteria, our main target sample consists of the 54 Very 
Large Array L-band calibrators which are unresolved in at least the Very Large 
Array (VLA) B-array (i.e. with angular extents $< 5''$) and which have L-band 
flux densities greater than 3~Jy. The flux density threshold of 3~Jy was used 
to ensure that the required dynamic range could be achieved in ``reasonable'' 
integration times ($\sim 10$~hours). In addition to the above 54 calibrators, 
we also included a few other compact sources that are unresolved by the 
shorter baselines of the WSRT. The GMRT and WSRT were chosen for the 
observations due to their high sensitivity and the high achievable spectral 
dynamic range \citep[e.g.][]{braun05,roy07}. In this paper, we present the 
spectra of a sub-sample of 32~sources, which have L-band flux densities 
greater than 3.5~Jy. 11 targets were observed with the GMRT and 23 with the 
WSRT, with two sources common to both telescopes, to cross-check the 
instrumental behaviour. Fig.~\ref{fig:obs} shows the position of these 32 
sources on the sky, in the Galactic coordinate system.

\subsection{The observations}
\label{sec:obs}

\begin{table*}
\caption{Summary of observational results}
\begin{center}
\begin{tabular}{lcccccccccc}
 \hline
Source & Coordinates & $S_{\rm 1.4}^a$ & $\tau_{\rm rms}$~$^b$ & $\tau_{\rm peak}$ & $\int\tau {\rm dV}$ & N(\hi)$^c$ & N(\hi,ISO)$^d$ & $\Ts\:\:^e$ & $\Delta V_{\rm 90}^{em~~f}$ & $\Delta V_{\rm 90}^{abs~~f}$ \\
& $(l,b)$ & Jy & $\times 10^{-3}$ & & \kms\ & $\times 10^{20}$~\cm & $\times 10^{20}$~\cm & K & \kms\ & \kms \\
\hline
\multicolumn{2}{l}{GMRT targets:}\\
\cline{1-1}
B0316$+$162     & 166.6,$-$33.6 & ~7.8 & 0.75 & $~~0.496$ & $~2.964 \pm 0.004$ & $~9.3$ &  $ 10.09 \pm 0.62  $ & $187 \pm 11$   & $~24$ & $ 12 $ \\
B0438$-$436     & 248.4,$-$41.6 & ~5.0 & 1.46 & $<0.0009$ & $~~~~~< 0.02~~~~~$ & $~1.3$ &  $ 1.319 \pm 0.012 $ & $ > 3618$      & $~63$ & $ - $ \\
B0531$+$194     & 186.8,~$-$7.1 & ~6.8 & 0.99 & $~~0.631$ & $~4.062 \pm 0.005$ & $26.5$ &  $ 28.8  \pm 1.5   $ & $ 389 \pm 21$  & $~32$ & $ 13 $ \\
B0834$-$196$^g$ & 243.3,$+$12.6 & ~5.0 & 1.08 & $~~0.187$ & $~0.973 \pm 0.005$ & $~6.8$ &  $ 6.86  \pm 0.21  $ & $ 387 \pm 12$  & $~68$ & $ 14 $ \\
B1151$-$348     & 289.9,$+$26.3 & ~5.0 & 1.05 & $~~0.120$ & $~0.714 \pm 0.004$ & $~7.0$ &  $ 7.06  \pm 0.12  $ & $ 542 \pm 10$  & $125$ & $ 23 $ \\
B1245$-$197     & 302.0,$+$42.9 & ~5.3 & 1.23 & $~~0.032$ & $~0.158 \pm 0.005$ & $~3.7$ &  $ 3.711 \pm 0.017 $ & $ 1288 \pm 41$ & $~55$ & $ 12 $ \\
B1345$+$125     & 347.2,$+$70.2 & ~5.2 & 1.07 & $~~0.086$ & $~0.305 \pm 0.005$ & $~1.9$ &  $ 1.920 \pm 0.011 $ & $ 345 \pm 6$   & $~34$ & $  7 $ \\
B1827$-$360     & 358.3,$-$11.8 & ~6.9 & 0.89 & $~~0.227$ & $~1.542 \pm 0.003$ & $~7.9$ &  $ 8.13  \pm 0.40  $ & $ 289 \pm 14 $ & $~58$ & $ 13 $ \\
B1921$-$293     & ~~9.3,$-$19.6 & ~6.0 & 1.02 & $~~0.377$ & $~1.446 \pm 0.006$ & $~7.1$ &  $ 7.44  \pm 0.41  $ & $ 282 \pm 16 $ & $~53$ & $ 12 $ \\
B2223$-$052     & ~59.0,$-$48.8 & ~5.7 & 0.79 & $~~0.148$ & $~1.034 \pm 0.003$ & $~4.5$ &  $ 4.61  \pm 0.22  $ & $ 245 \pm 12$ & $~23$ &  $ 16 $ \\
\hline                                                                                                                       
\multicolumn{2}{l}{WSRT targets:}\\                                                          
\cline{1-1}                                                                                  
B0023$-$263     &  42.3,$-$84.2 &  7.5 & 1.04 & $~0.0037$ & $0.0254 \pm 0.0050$ & $~1.6$ & $  1.623 \pm 0.013 $ & $3505 \pm 691$ & $101$ & $66$ \\
B0114$-$211     & 167.1,$-$81.5 &  3.7 & 1.58 & $~~0.044$ & $0.1354 \pm 0.0054$ & $~1.4$ & $  1.421 \pm 0.019 $ & $ 576 \pm  24$ & $~61$ & $8 $ \\
B0117$-$155     & 154.2,$-$76.4 &  4.2 & 1.51 & $~0.0067$ & $0.0307 \pm 0.0042$ & $~1.5$ & $  1.452 \pm 0.017 $ & $2594 \pm 356$ & $~36$ & $27$ \\
B0134$+$329$^g$ & 134.0,$-$28.7 & 16.5 & 0.53 & $~~0.058$ & $~0.443 \pm 0.002$  & $~4.3$ & $  4.330 \pm 0.019 $ & $ 536 \pm   3$ & $~43$ & $21$ \\ 
B0202$+$149     & 147.9,$-$44.0 &  3.5 & 0.98 & $~~0.084$ & $0.7472 \pm 0.0047$ & $~4.8$ & $  4.803 \pm 0.016 $ & $ 353 \pm   3$ & $~24$ & $13$ \\
B0237$-$233     & 209.8,$-$65.1 &  5.7 & 1.19 & $~~0.116$ & $0.2938 \pm 0.0040$ & $~2.1$ & $  2.132 \pm 0.020 $ & $ 398 \pm   7$ & $~32$ & $14$ \\
B0316$+$413     & 150.6,$-$13.3 & 23.9 & 0.49 & $~~0.230$ & $~1.941 \pm 0.003$  & $13.2$ & $ 13.64  \pm 0.51  $ & $ 385 \pm  14$ & $~49$ & $28$ \\
B0355$+$508     & 150.4,~$-$1.6 & ~5.3 & 2.12 & $~~6.438$ & $45.8 \pm 1.1$      & $74.4$ & $ 115    \pm 43    $ & $ 138 \pm  52$ & $~59$ & $36$ \\
B0404$+$768     & 133.4,$+$18.3 & ~5.8 & 1.16 & $~~0.424$ & $~1.945 \pm 0.005$  & $10.8$ & $ 11.06  \pm 0.42  $ & $ 312 \pm  12$ & $124$ & $17$ \\
B0429$+$415     & 161.0,~$-$4.3 & ~8.6 & 0.90 & $~~0.716$ & $10.879 \pm 0.007$  & $37.1$ & $ 41.3  \pm 1.6    $ & $ 208 \pm   8$ & $~66$ & $66$ \\
B0518$+$165     & 187.4,$-$11.3 & ~8.5 & 1.08 & $~~1.130$ & $~6.241 \pm 0.007$  & $20.7$ & $ 23.6  \pm 5.4    $ & $ 207 \pm  47$ & $~43$ & $17$ \\
B0538$+$498     & 161.7,$+$10.3 & 22.5 & 0.49 & $~~0.912$ & $~5.618 \pm 0.003$  & $19.5$ & $ 21.31  \pm 0.94  $ & $ 208 \pm   9$ & $~71$ & $18$ \\
B0831$+$557     & 162.2,$+$36.6 & ~8.8 & 1.15 & $~~0.089$ & $~0.483 \pm 0.005$  & $~4.5$ & $ 4.499  \pm 0.020 $ & $ 511 \pm   6$ & $~75$ & $60$ \\
B0906$+$430     & 178.3,$+$42.8 &  3.9 & 0.82 & $~0.0059$ & $0.0512 \pm 0.0030$ & $~1.3$ & $  1.283 \pm 0.020 $ & $1375 \pm  83$ & $~99$ & $29$ \\
B1328$+$254     &  22.5,$+$81.0 &  6.8 & 0.33 & $~0.0016$ & $0.0214 \pm 0.0031$ & $~1.1$ & $  1.108 \pm 0.019 $ & $2840 \pm 414$ & $~67$ & $67$ \\
B1328$+$307     &  56.5,$+$80.7 & 14.7 & 0.30 & $~0.0082$ & $0.0717 \pm 0.0018$ & $~1.2$ & $  1.221 \pm 0.016 $ & $ 934 \pm  26$ & $~54$ & $28$ \\
B1611$+$343     &  55.2,$+$46.4 &  4.8 & 0.72 & $~0.0033$ & $0.0187 \pm 0.0027$ & $~1.4$ & $  1.360 \pm 0.019 $ & $3989 \pm 579$ & $~66$ & $31$ \\
B1641$+$399     &  63.5,$+$40.9 &  8.9 & 0.49 & $~0.0014$ & $0.0088 \pm 0.0024$ & $~1.1$ & $  1.078 \pm 0.018 $ & $6720 \pm1836$ & $~75$ & $81$ \\
B2050$+$364     & ~78.9,~$-$5.1 & ~5.2 & 1.46 & $~~0.351$ & $~3.024 \pm 0.010$  & $27.7$ & $ 28.53  \pm 0.96  $ & $ 518 \pm  17$ & $~87$ & $25$ \\
B2200$+$420     & ~92.6,$-$10.4 & ~6.1 & 2.68 & $~~0.571$ & $~3.567 \pm 0.016$  & $17.1$ & $ 18.59  \pm 0.98  $ & $ 286 \pm  15$ & $~91$ & $18$ \\
B2203$-$188     &  36.7,$-$51.2 &  6.0 & 1.00 & $~~0.066$ & $0.2483 \pm 0.0042$ & $~2.4$ & $ 2.381  \pm 0.011 $ & $ 526 \pm   9$ & $~48$ & $25$ \\
B2348$+$643     & 116.5,~$+$2.6 & ~5.0 & 1.31 & $~~2.713$ & $32.517 \pm 0.031$ & $70.2$  & $ 88     \pm 16    $ & $ 148 \pm  27$ & $103$ & $99$ \\
\hline                                                                                                                                                    
\multicolumn{2}{l}{ATCA sources:}\\                                                                                                                   
\cline{1-1}                                                                                                                                          
B0407$-$658     & 278.6,$-$40.9 &  16.2 & 1.0 & $~0.138$ & $0.5481 \pm 0.0070$ & $~3.4$  & $  3.408 \pm 0.030$  & $341 \pm 5 $   & $64$ & $31$ \\
B1814$-$637     & 330.9,$-$20.8 &  14.4 & 1.0 & $~0.374$ & $0.9974 \pm 0.0067$ & $~6.5$  & $  6.66 \pm 0.27 $   & $366 \pm 15$   & $39$ & $~6$ \\
\end{tabular}\\                                                                                                                                       
\end{center}                                                                                                                                         
\begin{flushleft}                                                                                                                                   
$^a$~The quoted 1.4~GHz flux densities are from the VLA calibrator manual, 
except for B0407$-$658 and B1814$-$637, which are from \citet{kanekar03b}. \\
$^b$~The ``off-line'' RMS optical depth values of column~(4) are at velocity 
resolutions of $\approx 0.4$~\kms\ for all GMRT and ATCA spectra and $\approx 
0.26$~\kms\ for all WSRT spectra, except for B1328+254 and B1328+307 ($\approx 
0.52$~\kms). \\
$^c$~The \hi\ column density N(\hi) assumes the optically-thin limit, using 
the brightness temperatures measured in the LAB survey. These should be 
treated as lower limits to the true \hi\ column density. \\
$^d$~The \hi\ column density N(\hi, ISO) from the isothermal estimate, 
N(\hi)~$= 1.823 \times \int \tau \Tb/ [1 - e^{-\tau}] dV$ \citep{chengalur13}. 
The error on this estimate by adding in quadrature the errors on the 
N(\hi,ISO) values of individual velocity channels, with these errors in turn 
obtained from the Monte Carlo simulations of \citet{chengalur13}. 
Specifically, for $\tau < 0.1$, we assumed the optically-thin limit (i.e. 
N(\hi)~$= 1.823 \times \int \Tb dV$), with standard error propagation on the 
errors on the measured brightness temperature. For $0.1 \le \tau < 1$, we 
assumed 30\% errors on the inferred N(\hi,ISO) value, while for $\tau > 1$, we 
assumed errors of a factor of 3 on the inferred N(\hi,ISO) value. Note that 
very few velocity channels in the sample have $\tau > 1$, due to which the 
final errors on the N(\hi,ISO) estimate are relatively low, $< 30$\% in all 
cases and typically much lower.\\
$^e$~The column-density-weighted harmonic mean spin temperature $\Ts$ along 
the sightline, derived from N(\hi,ISO) and $\int\tau{\rm dV}$.\\ 
$^f$~The velocity range containing 90\% of the integrated \hii\ emission or 
optical depth.\\
$^g$~For sources observed with both the telescopes, the WSRT spectrum for 
B0134$+$329 and the GMRT spectrum for B0834$-$196 have been used for the later 
analysis due to their higher signal-to-noise ratio.
\end{flushleft}
\label{table:results}
\end{table*}

The observations were aimed at detecting wide absorption from the WNM (of 
width $\gtrsim 20$~\kms) as well as resolving out narrow CNM absorption 
features. This required both large bandwidths ($> 100$~\kms) and high spectral 
resolution ($\lesssim 0.5$~\kms). The GMRT observations used a single 
intermediate frequency (IF) band and the hardware correlator, with two 
polarizations and a baseband bandwidth of 0.5~MHz, sub-divided into 256 
channels. This yielded a total velocity coverage of $\sim 105$~\kms\ and a 
velocity resolution of $\sim 0.4$~\kms. Conversely, the WSRT observations used 
two IF bands, two polarizations and the DZB correlator, with each baseband 
having a bandwidth of 2.5~MHz (i.e. a velocity coverage of $\sim 525$~\kms). 
For 20 sources, the 2.5~MHz baseband was sub-divided into 2048 channels, 
giving a resolution of $\sim 0.26$~\kms; for two sources (B1328+254 and 
B1328+307), 1024 channels were used, implying a velocity resolution of $\sim 
0.52$~\kms.

Since Galactic \hii\ absorption is likely to be present in all directions 
(i.e. towards all possible bandpass calibrators), bandpass calibration was 
done by frequency-switching. Bandpass calibration for the GMRT observations 
was carried out by frequency-switching at the first local oscillator, using a 
frequency throw of 5~MHz every 5~minutes, with a duty cycle of $50\%$. For the 
WSRT observations, bandpass calibration was done with in-band 
frequency-switching, with a frequency throw of 1~MHz, again every 5~minutes. 
The central frequencies of the two IF bands were carefully chosen to ensure 
that the \hii\ absorption line was always within the 2.5~MHz observing band, 
thus maximizing observing efficiency. The total observing time for each 
source, including all calibration and overheads, was 10~hours with the GMRT 
and 12~hours at the WSRT. As noted above, the integration time was chosen to 
be sufficient to detect absorption from warm gas  with $\Tk \sim 8000$\,K 
\citep[${\rm T_s} \sim 5000$\,K;][]{liszt01}, \hi\ column density $\sim 2 
\times 10^{20}$\,\cm, and line FWHM similar to the thermal width. A summary of 
the observational details is given in Table~\ref{table:obs}.

\subsection{Data analysis}
\label{sec:data}

All data were analysed in the Astronomical Image Processing System of the 
National Radio Astronomy Observatory (NRAO {\sc AIPS}), using standard 
procedures. Some of these data were edited out for a variety of reasons (e.g. 
dead antennas, correlator problems, radio frequency interference, shadowing, 
etc). After removing all such bad data, standard calibration procedures were 
followed to obtain the antenna-based complex gains. Bandpass calibration was 
applied using interpolation between the bandpass solutions (i.e. {\small 
DOBAND}=3 in {\sc AIPS}). In all cases, the background continuum source was 
unresolved or barely resolved, with all the flux in a single compact 
component. For all sources, a linear fit to line-free channels on each 
interferometer baseline was then used to subtract out the continuum, using the 
task {\sc UVLIN} in {\sc AIPS}). The task {\sc CVEL} was then used to shift 
the residual visibility data to the local standard of rest (LSR) frame. The 
residual U-V data were then imaged in all channels, and the final flux density 
spectrum obtained by a cut through the dirty cube at the location of the 
target source. This spectrum was then converted into optical depth units, 
using the flux density estimated from the continuum image of the line-free 
channels. 

\hii\ absorption was detected towards every source except one, B0438$-$436; 
the final optical depth spectra for all sources are shown in the middle panel 
of Fig.~\ref{fig:spectra}. For completeness, this figure also includes the two 
sources (B0407$-$658 and B1814$-$637) observed with the ATCA by 
\citet{kanekar03b}. For the two sources (B0134$+$329 and B0834$-$196) that 
were observed with both GMRT and WSRT, we have carried out a comparison 
between the WSRT and GMRT spectra and find that they are in excellent 
agreement, within the noise. This indicates that there are no systematic 
effects in the absorption spectra, either due to instrumental artefacts or 
unresolved \hii\ emission. We will use the WSRT spectrum for B0134$+$329 and 
the GMRT spectrum for B0834$-$196 in the later analysis, as these have higher 
signal-to-noise ratios.

For each sightline, the Galactic \hii\ emission spectrum was obtained from the 
website of the LAB 
survey\footnote{http://www.astro.uni-bonn.de/en/download/data/lab-survey/} 
\citep{hartmann97,bajaja05,kalberla05}. These spectra have a velocity 
resolution of $\sim 1$~\kms, slightly worse than the resolutions of our GMRT 
and WSRT absorption spectra, and, for each source, are shown in the top panels 
of Fig.~\ref{fig:spectra}. The LAB emission spectra were used to derive the 
\hi\ column density along each sightline for two limiting cases. The first is 
the traditional optically-thin limit, i.e. assuming that the peak \hii\ 
optical depth along the sightline is $\ll 1$; this yields ${\rm T_B} \approx 
{\rm T_s} \times \tau$, allowing one to estimate the \hi\ column density 
directly from the measured brightness temperature ${\rm T_B}$. The low 
optical-depth assumption is a good one for 20 sources of the sample, where the 
peak optical depth $\tau_{\rm peak} \lesssim 0.2$. However, for 14 targets, 
the peak optical depth is $> 0.2$, implying that the optically-thin estimate 
of the \hi\ column density is a lower limit. We have hence also estimated the 
\hi\ column density using the ``isothermal estimate'' 
\citep{dickey82,chengalur13}. This gives a statistically unbiased estimate of 
the \hi\ column density by combining the measured brightness temperature and 
the total \hii\ optical depth \citep{chengalur13}, even without knowing the 
relative position of different clouds along the sightline or the optical depth 
of individual components. 

Finally, for the absorption spectra, it is important to emphasize that the 
system temperature is higher at the frequencies at which there is Galactic 
emission; the noise level hence shows significant variation with frequency. 
For bandpass calibration with frequency-switching, there are two contributions 
to the noise: (i)~the noise in the on-source spectrum $n_c$, and (ii)~the 
noise from the bandpass spectrum, $n_b$. It is the former that contains a 
contribution from the brightness temperature of the Galactic \hii\ emission 
and hence depends on frequency. Conversely, since the bandpass is measured at 
an offset frequency, $n_b$ does not change with frequency. We hence estimated 
the noise at each frequency channel in the following manner: For each optical 
depth spectrum, we first measure the ``off-line'' noise from a range of 
line-free channels. This noise has equal contributions from $n_c$ and $n_b$, 
and thus allows us to determine $n_{c,off}$, the noise due to the telescope 
system temperature on cold sky (${\rm T_{sys}} \sim 70$\,K for the GMRT and 
${\rm T_{sys}} \sim 30$\,K for the WSRT). We then determine $n_c$ for each 
channel by scaling $n_{c,off}$ by the factor $({\rm T_B} + {\rm T_{sys}})/{\rm 
T_{sys}}$, using the brightness temperature measured in the LAB survey. 
Finally, for each channel, we add $n_c$ and $n_b$ in quadrature to obtain the 
full optical depth noise spectrum. We note, in passing, that this implicitly 
assumes that the fields of view of the GMRT, the WSRT and the LAB telescopes 
are the same. While this assumption is not entirely true for the GMRT (whose 
telescopes have an effective dish diameter of 35\,m at 1.4~GHz, compared to 
the 25\,m dish diameter of the others), this is unlikely to have a significant 
effect on our results.

\setcounter{figure}{1}
\begin{figure*}
\begin{center}
\includegraphics[height=2.2in,angle=-90]{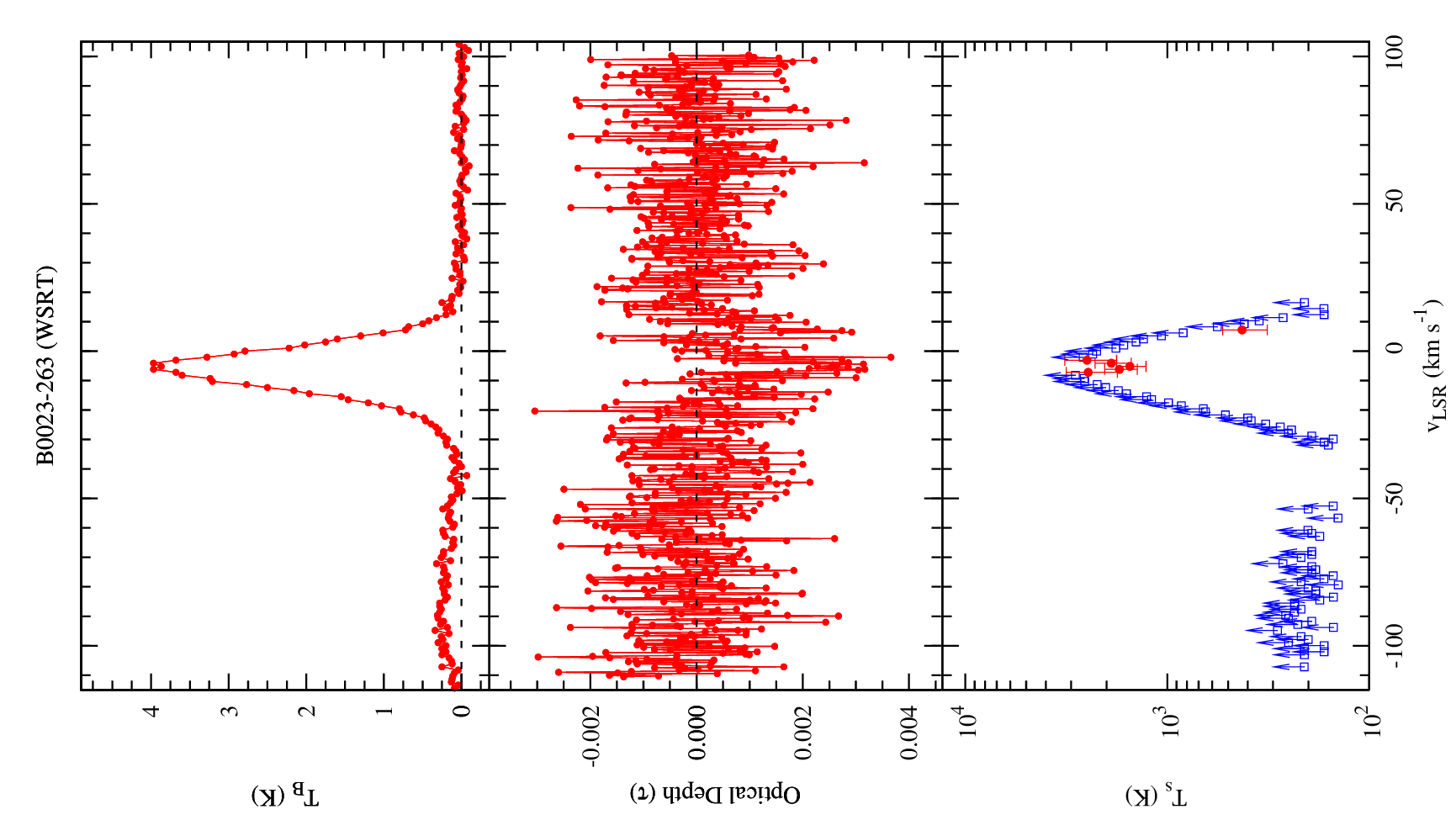}
\includegraphics[height=2.2in,angle=-90]{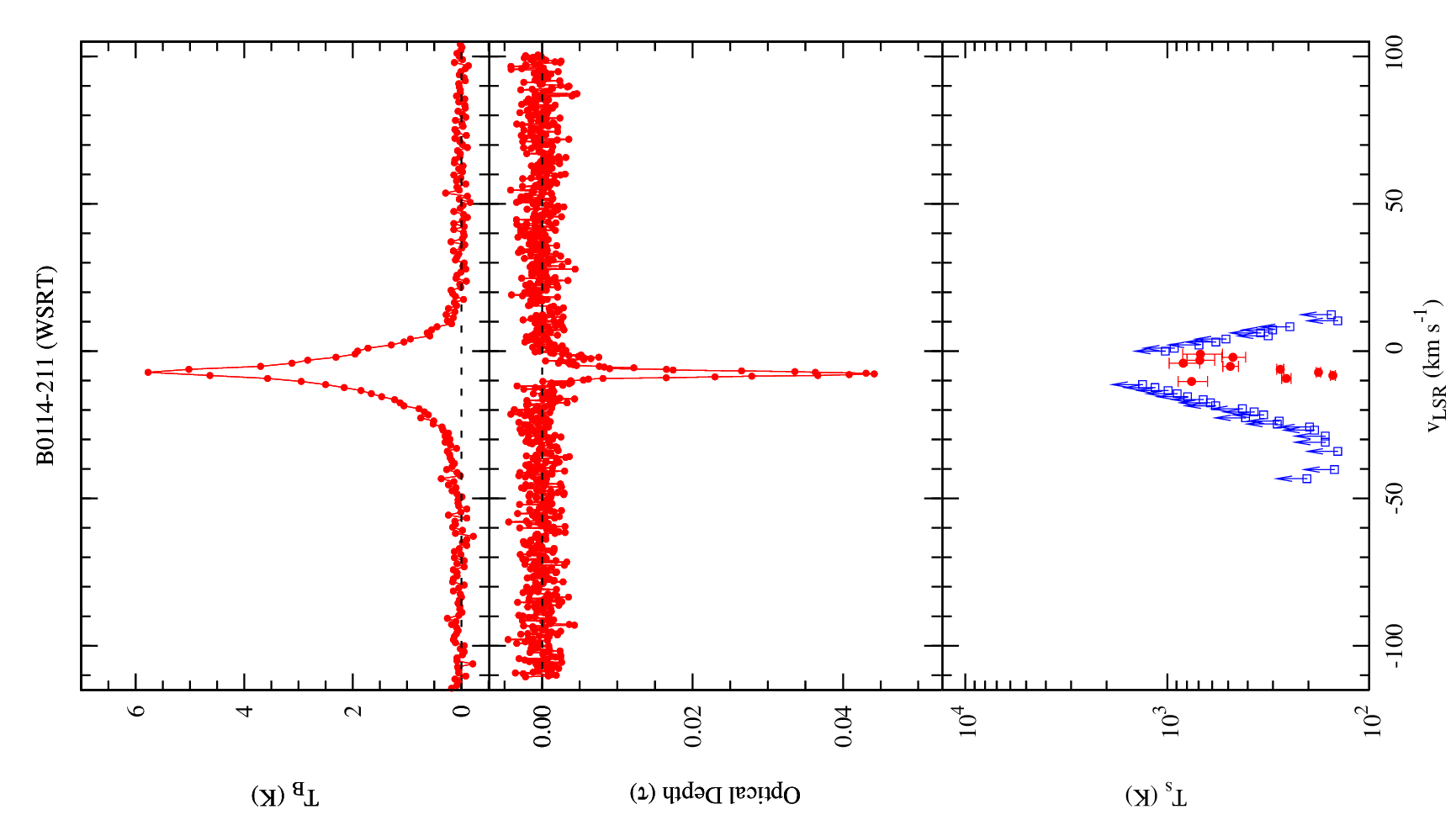}
\includegraphics[height=2.2in,angle=-90]{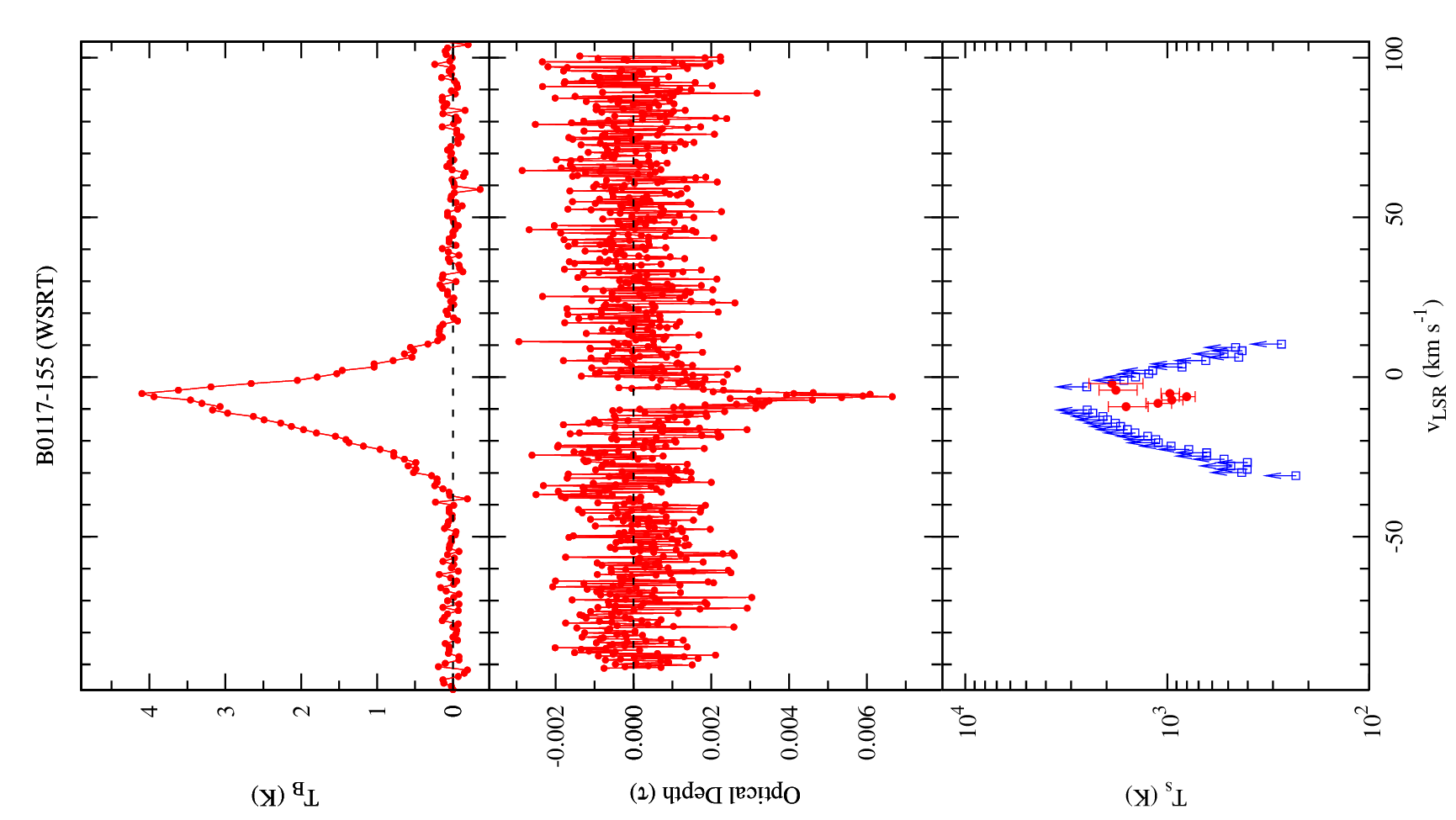}
\includegraphics[height=2.2in,angle=-90]{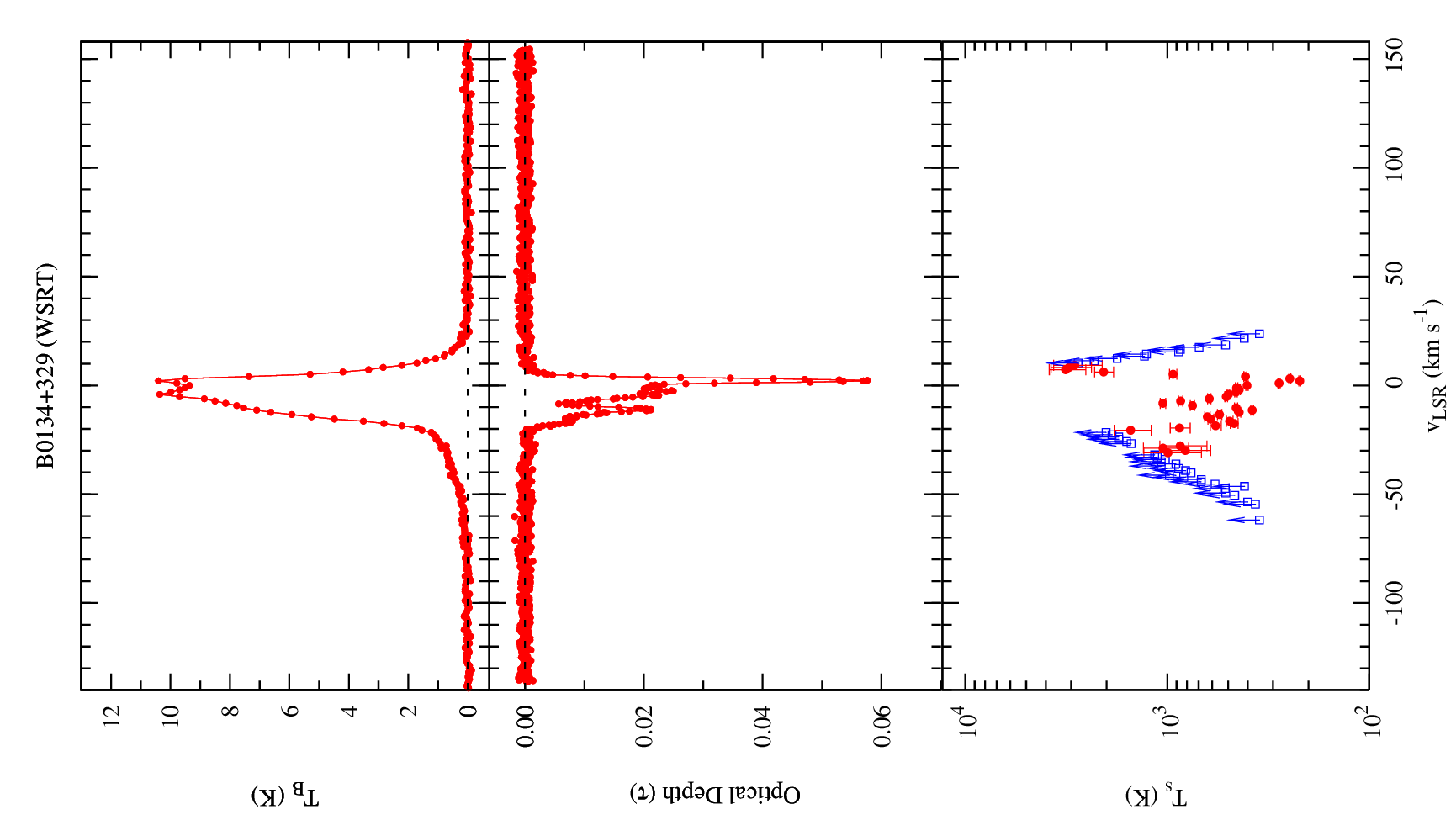}
\includegraphics[height=2.2in,angle=-90]{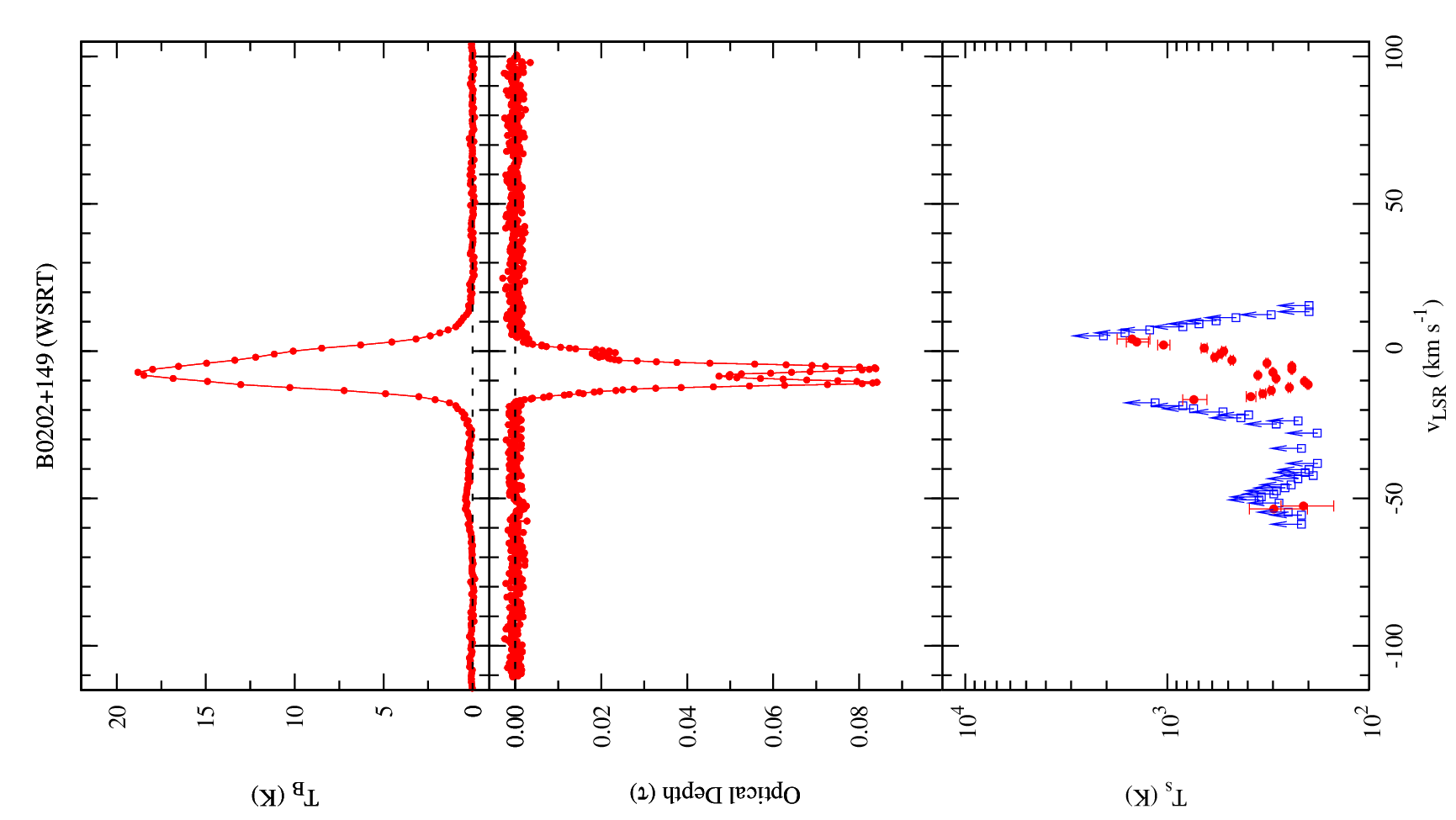}
\includegraphics[height=2.2in,angle=-90]{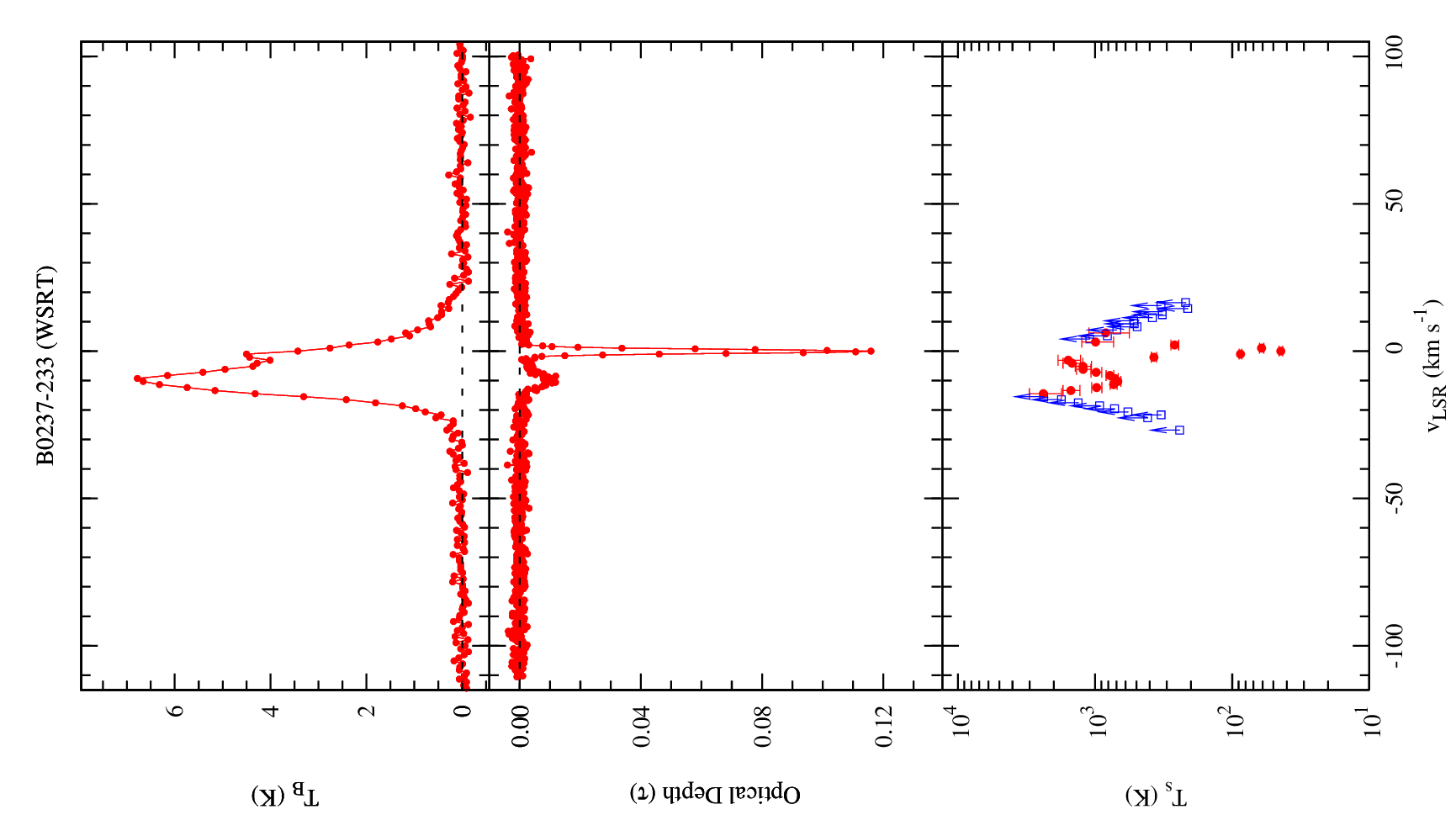}
\caption{\label{fig:spectra} \hii\ absorption spectra obtained using the GMRT, 
the WSRT or the ATCA. For each source, the top, middle and bottom panels 
contain, respectively, the \hii\ emission spectrum (from the LAB survey), the 
\hii\ absorption spectrum, and the spin temperature spectrum. The \hii\ 
absorption spectrum is at the original resolution (i.e. $\sim 0.26$ or $\sim 
0.52$~\kms\ for the WSRT spectra or $\sim 0.4$~\kms\ for the GMRT/ATCA 
spectra), while the emission and spin temperature spectra are at the velocity 
resolution of the LAB survey (i.e. 1.03~\kms). The name of the source and the 
telescope (GMRT/WSRT/ATCA) is mentioned above each figure. The plots are 
restricted to the LSR velocity range containing \hii\ absorption.}
\end{center}
\end{figure*}

\setcounter{figure}{1}
\begin{figure*}
\begin{center}
\includegraphics[height=2.2in,angle=-90]{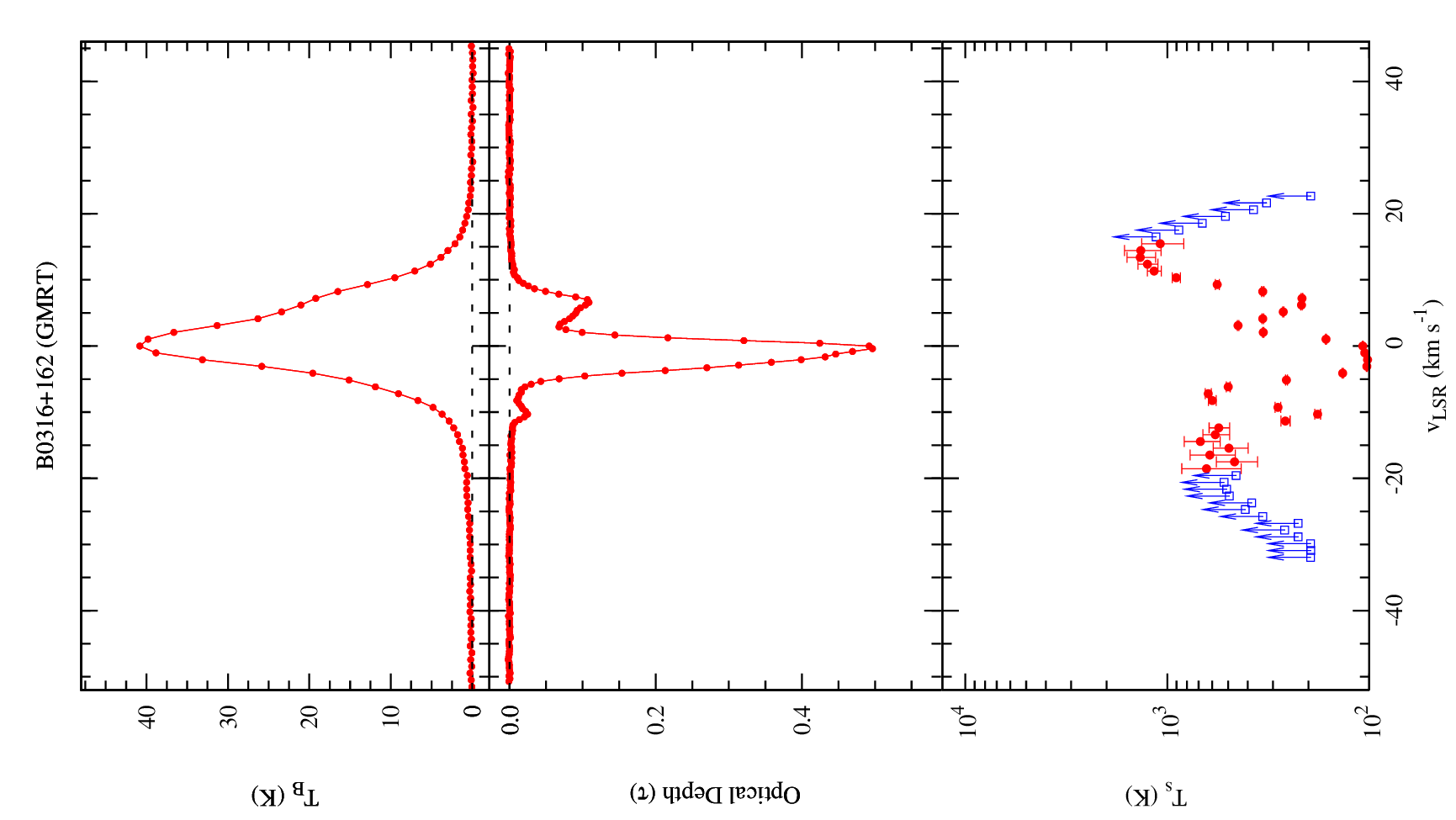}
\includegraphics[height=2.2in,angle=-90]{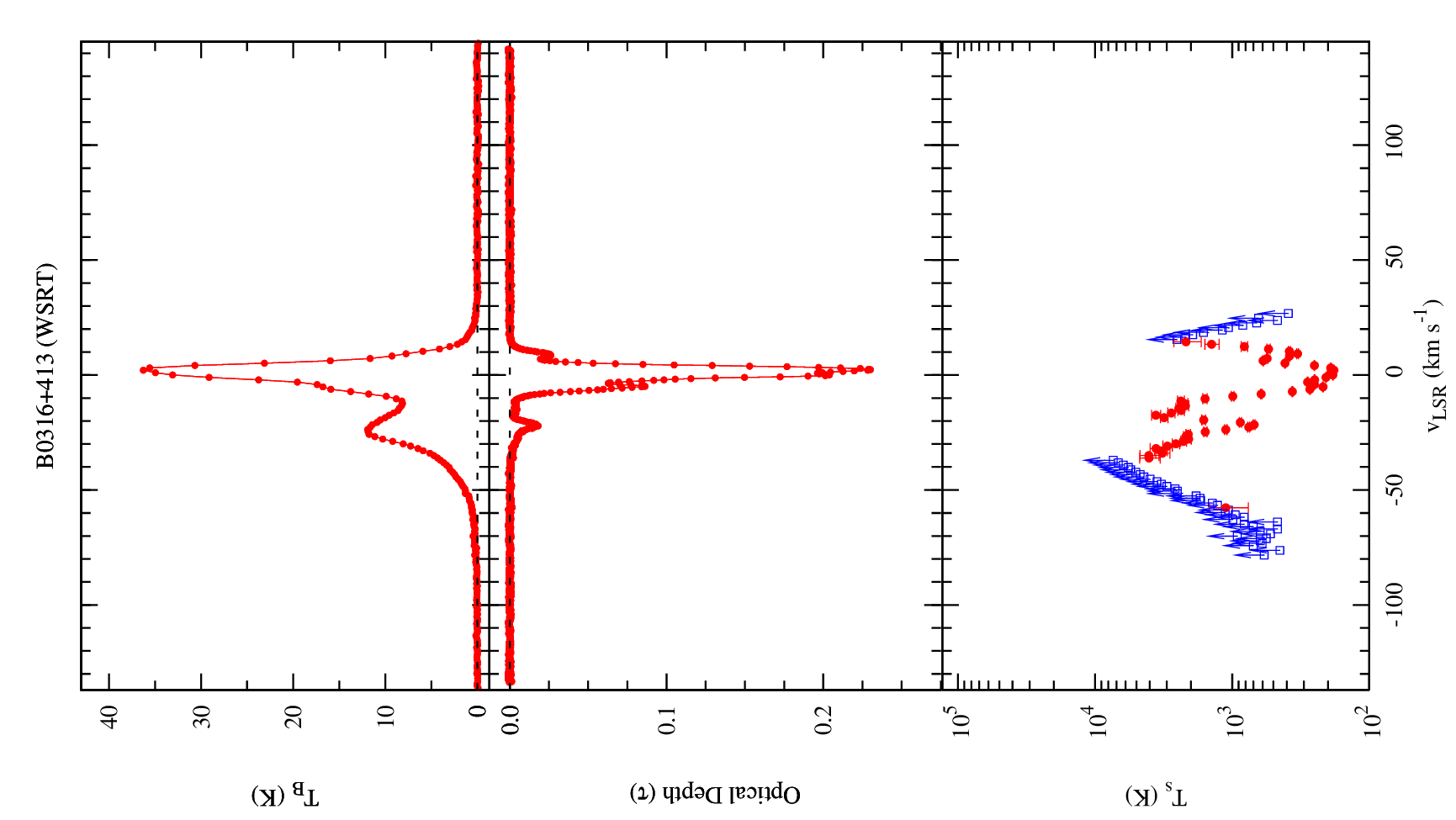}
\includegraphics[height=2.2in,angle=-90]{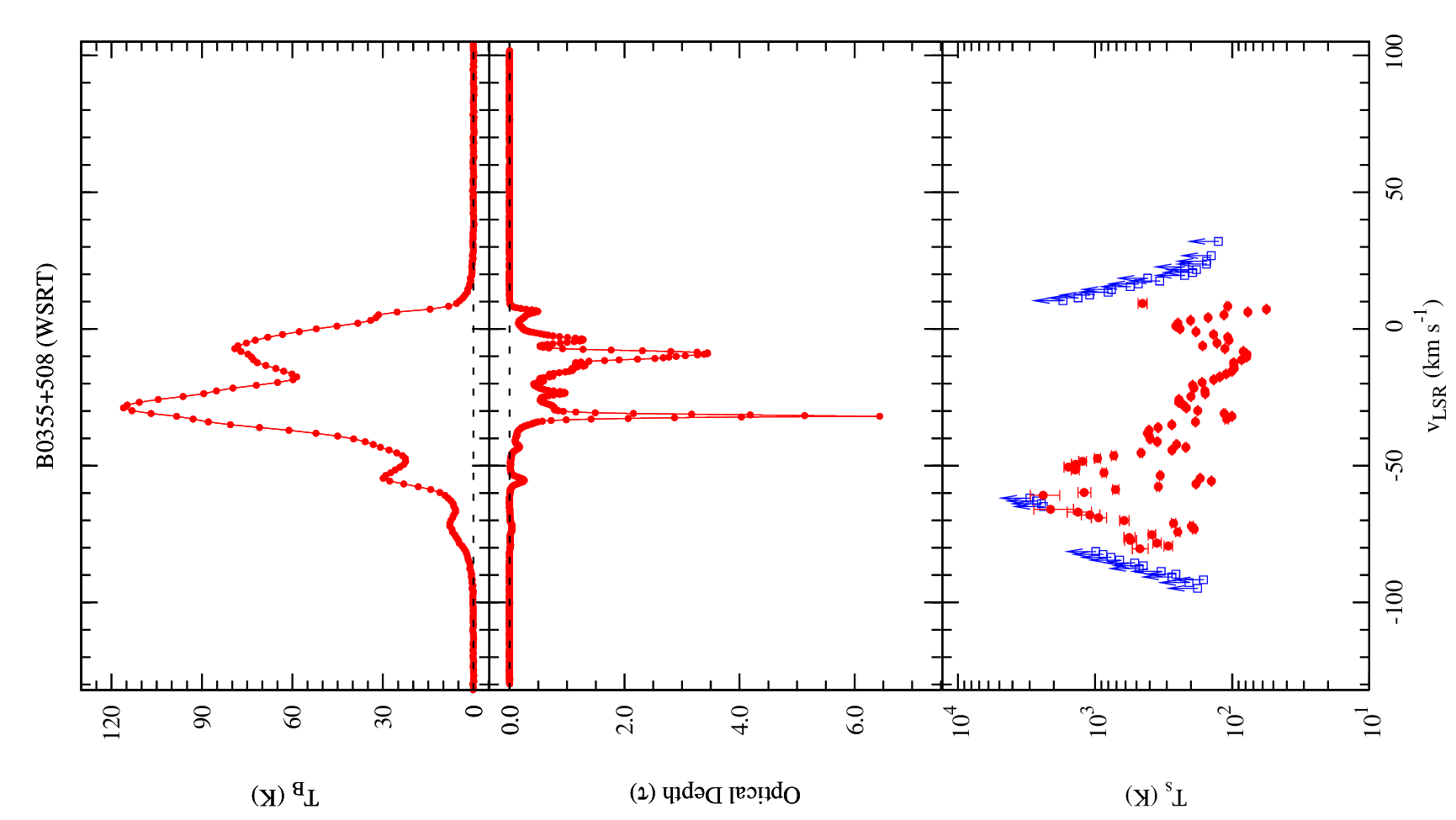}
\includegraphics[height=2.2in,angle=-90]{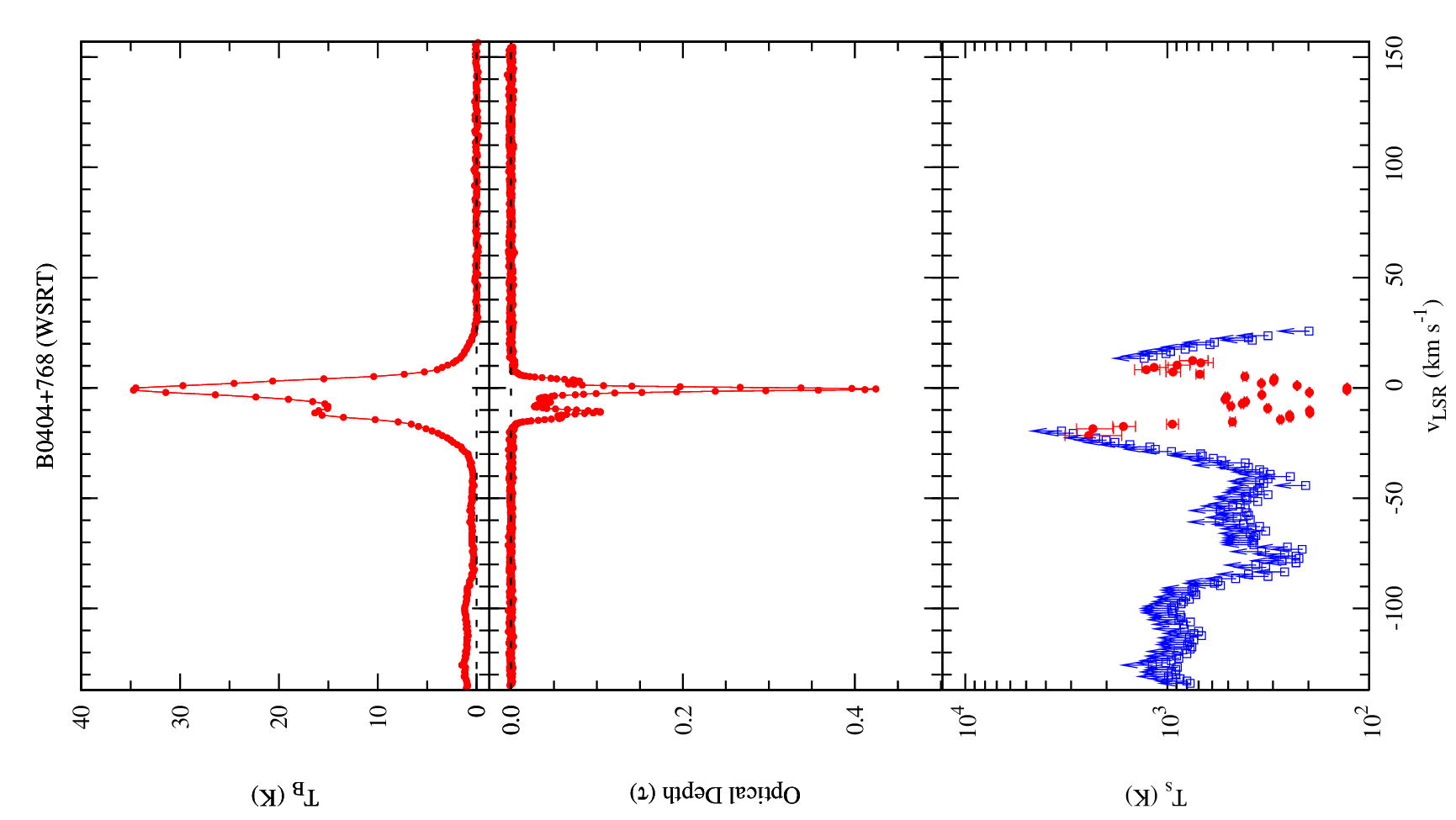}
\includegraphics[height=2.2in,angle=-90]{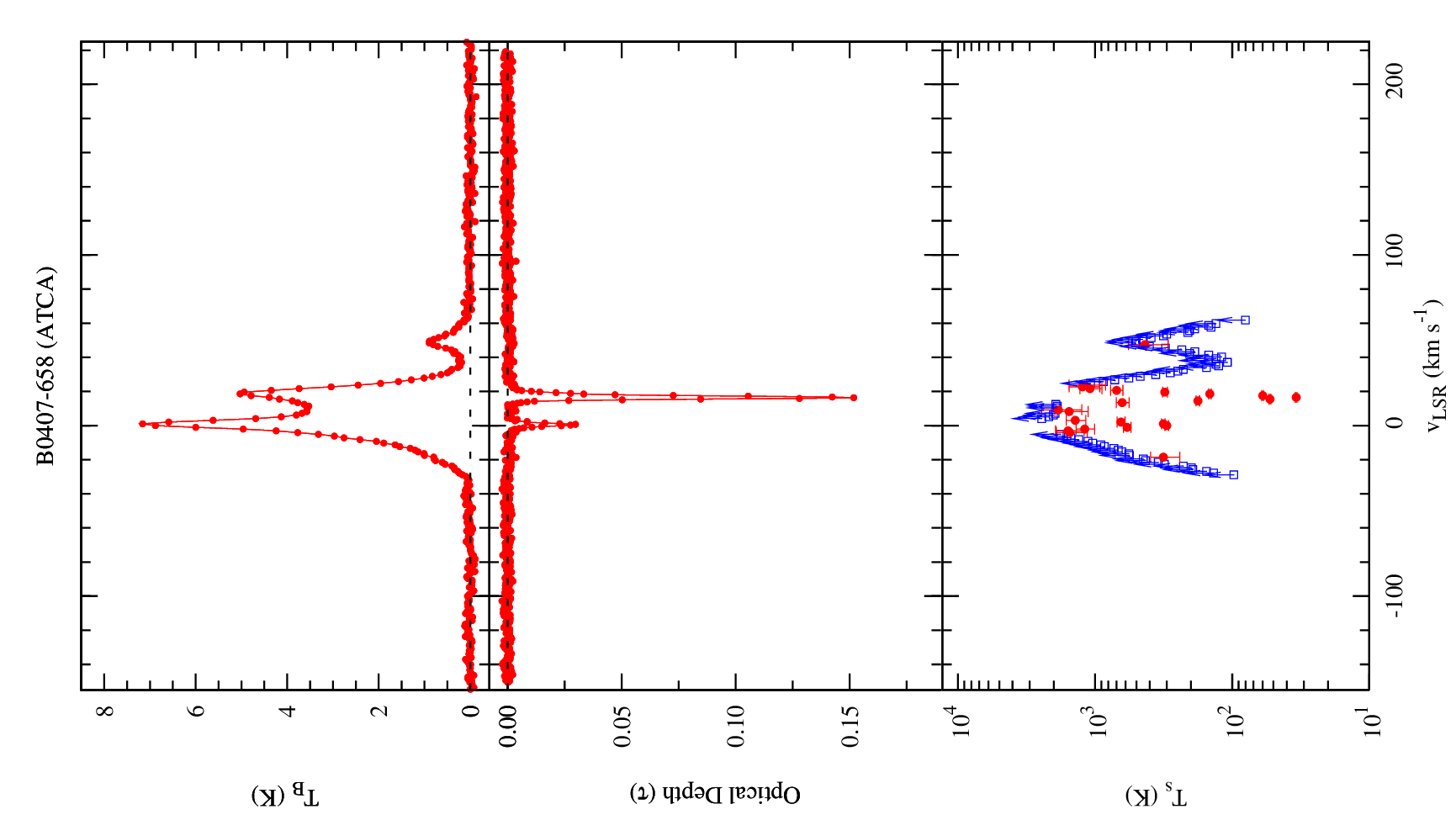}
\includegraphics[height=2.2in,angle=-90]{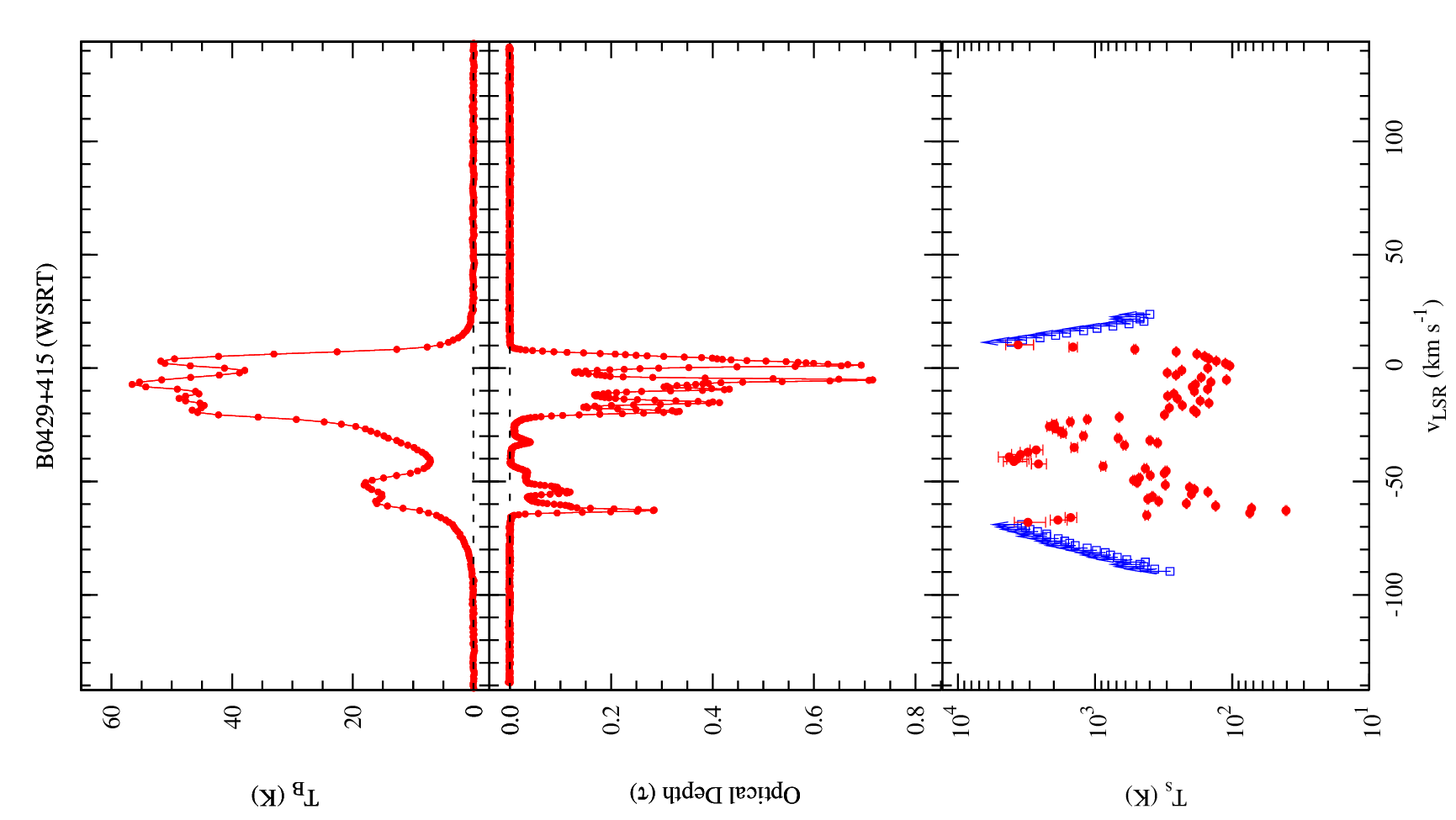}
\caption{\label{fig:spectra-2} ({\it continued}) \hii\ absorption spectra 
obtained using the GMRT/WSRT/ATCA, \hii\ emission spectra from the LAB survey 
and the spin temperature spectra.}
\end{center}
\end{figure*}

\setcounter{figure}{1}
\begin{figure*}
\begin{center}
\includegraphics[height=2.2in,angle=-90]{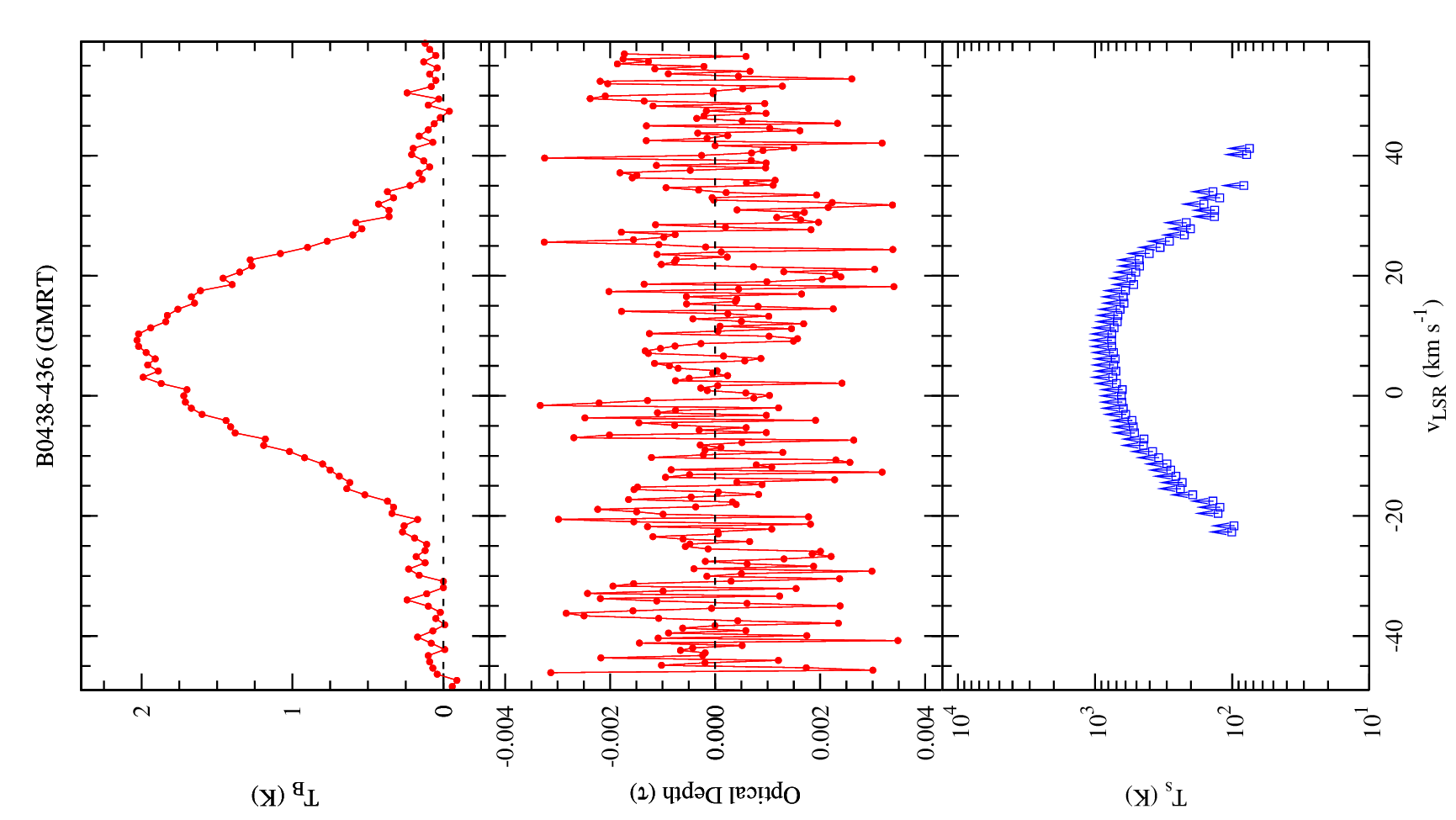}
\includegraphics[height=2.2in,angle=-90]{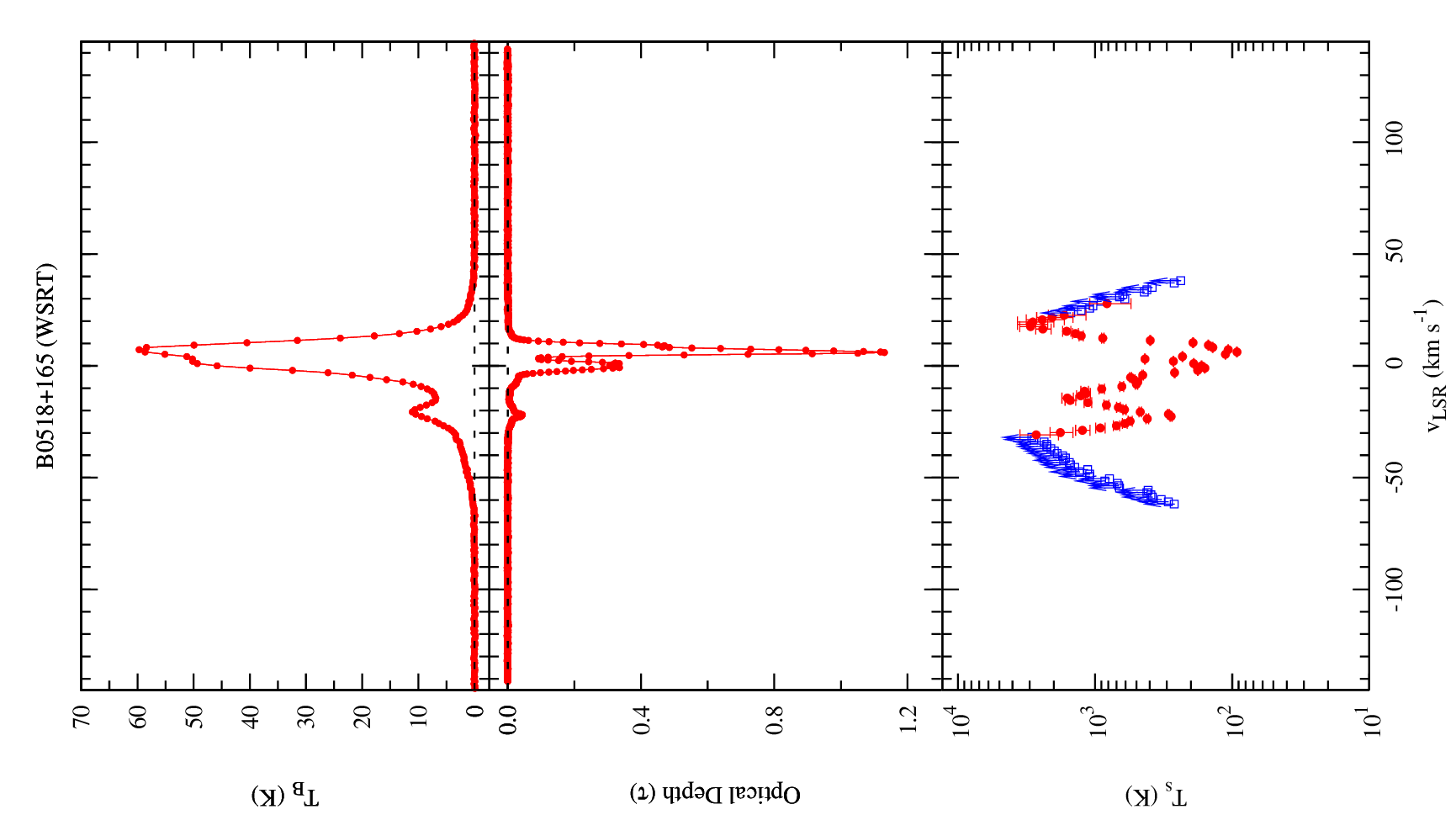}
\includegraphics[height=2.2in,angle=-90]{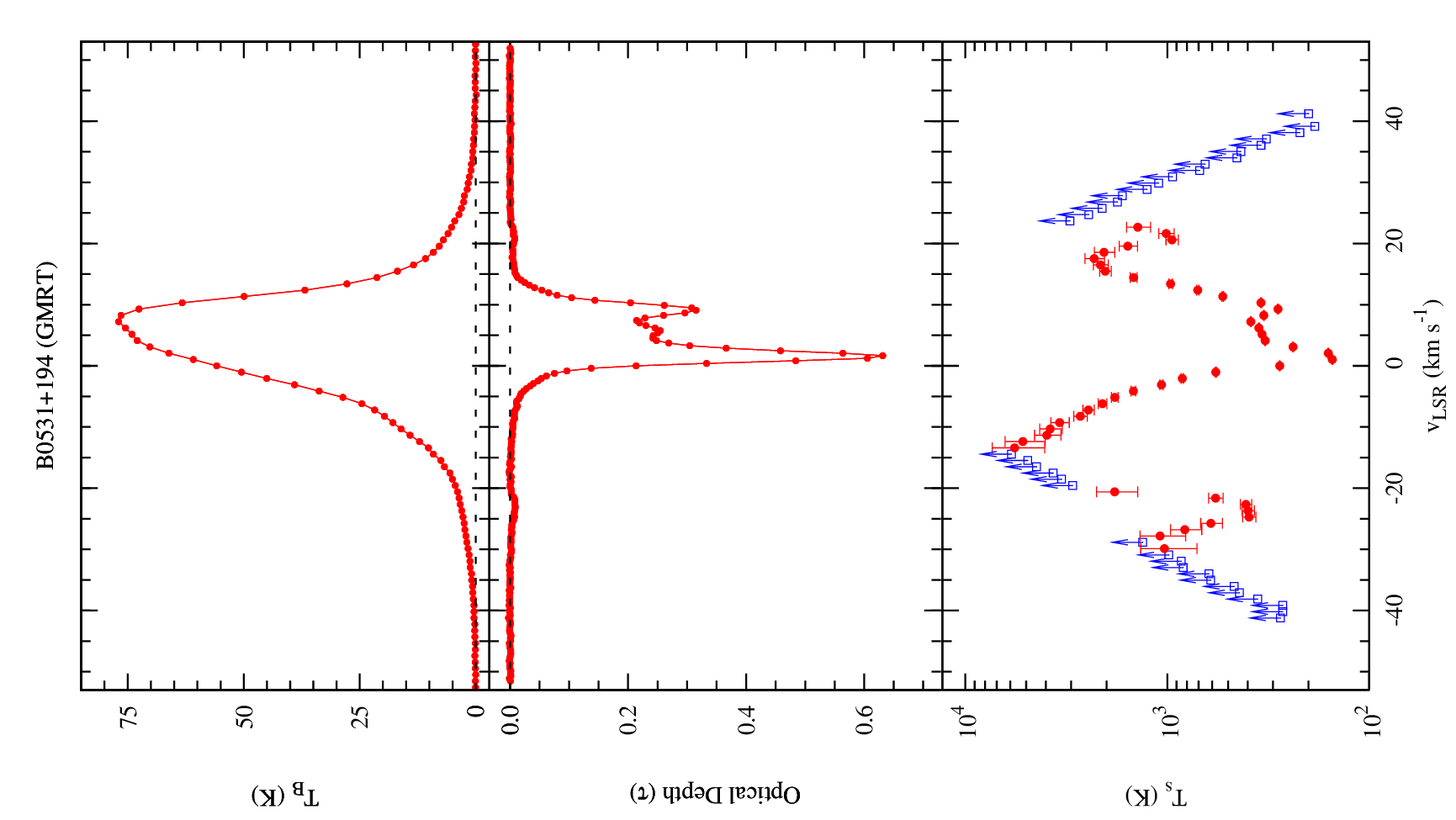}
\includegraphics[height=2.2in,angle=-90]{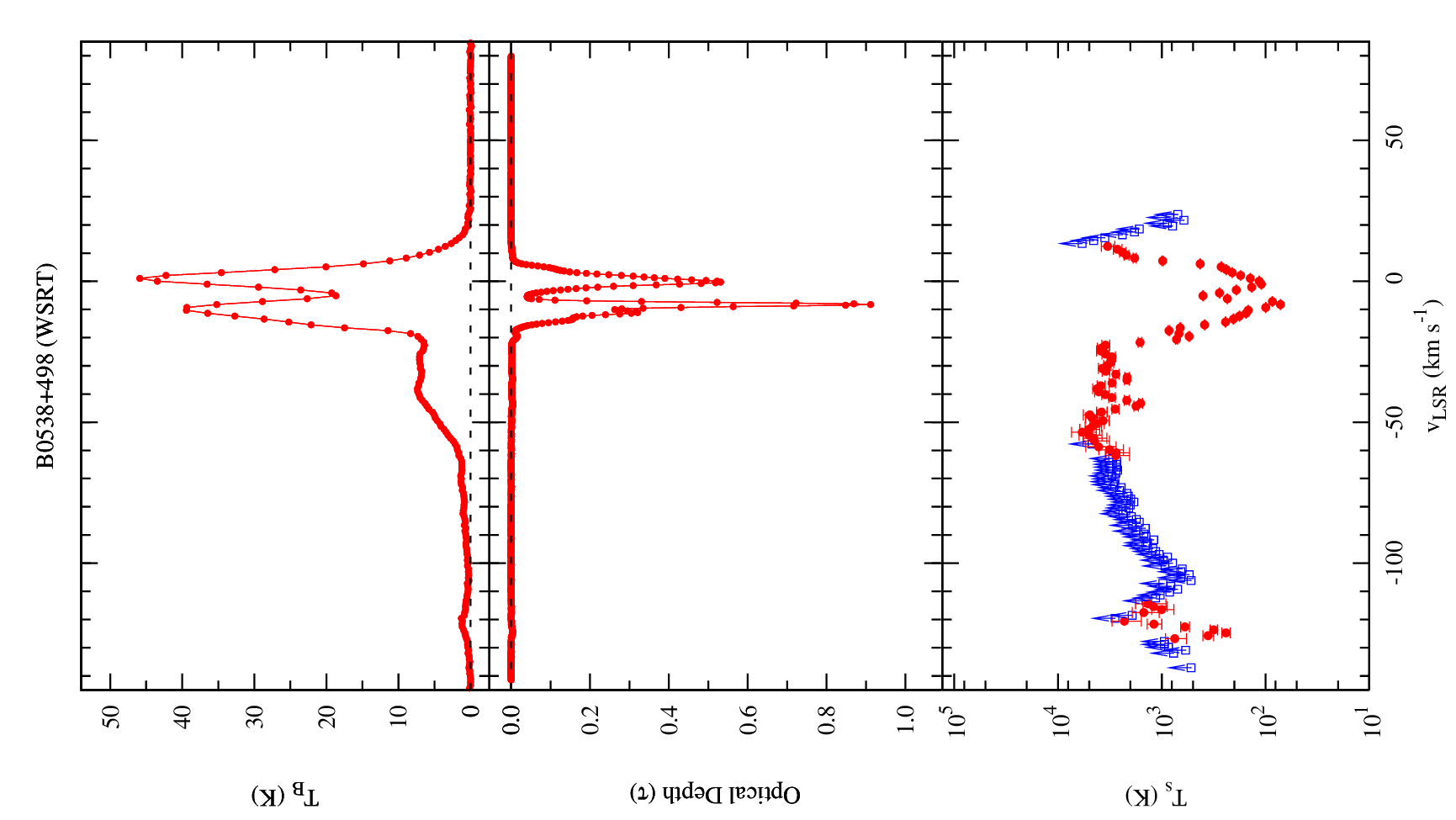}
\includegraphics[height=2.2in,angle=-90]{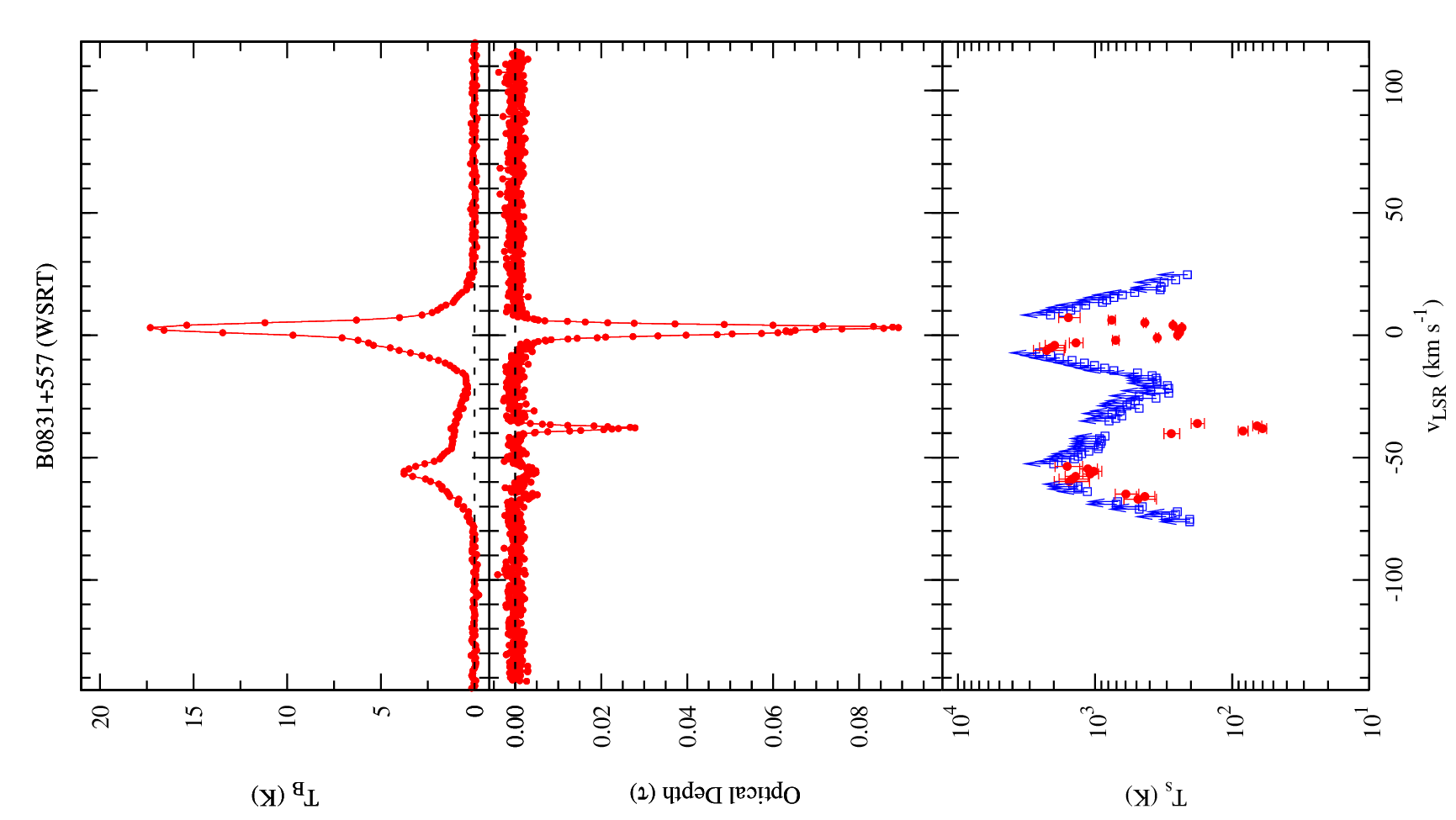}
\includegraphics[height=2.2in,angle=-90]{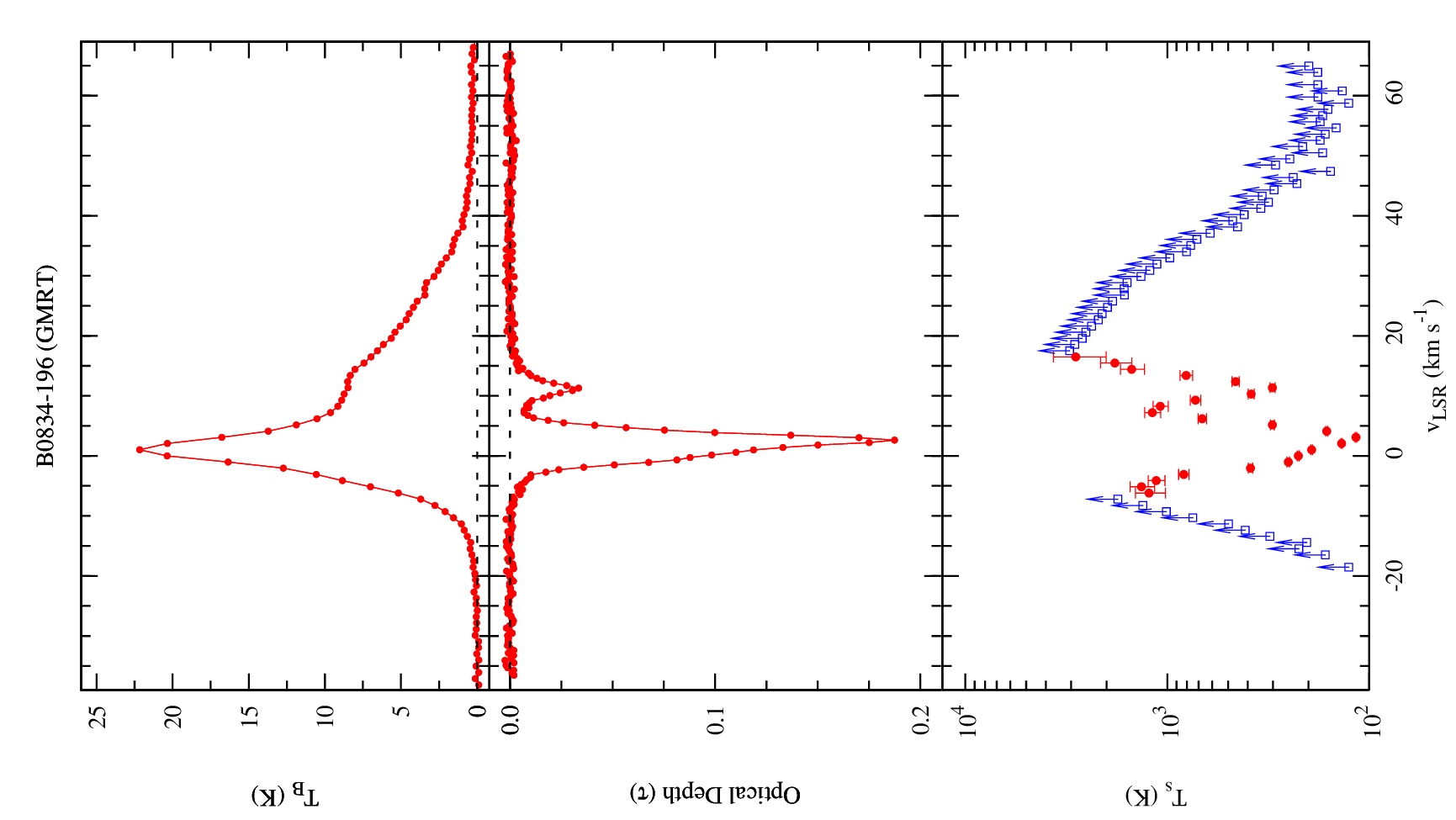}
\caption{\label{fig:spectra-3} ({\it continued}) \hii\ absorption spectra 
obtained using the GMRT/WSRT/ATCA, \hii\ emission spectra from the LAB survey 
and the spin temperature spectra.}
\end{center}
\end{figure*}

\setcounter{figure}{1}
\begin{figure*}
\begin{center}
\includegraphics[height=2.2in,angle=-90]{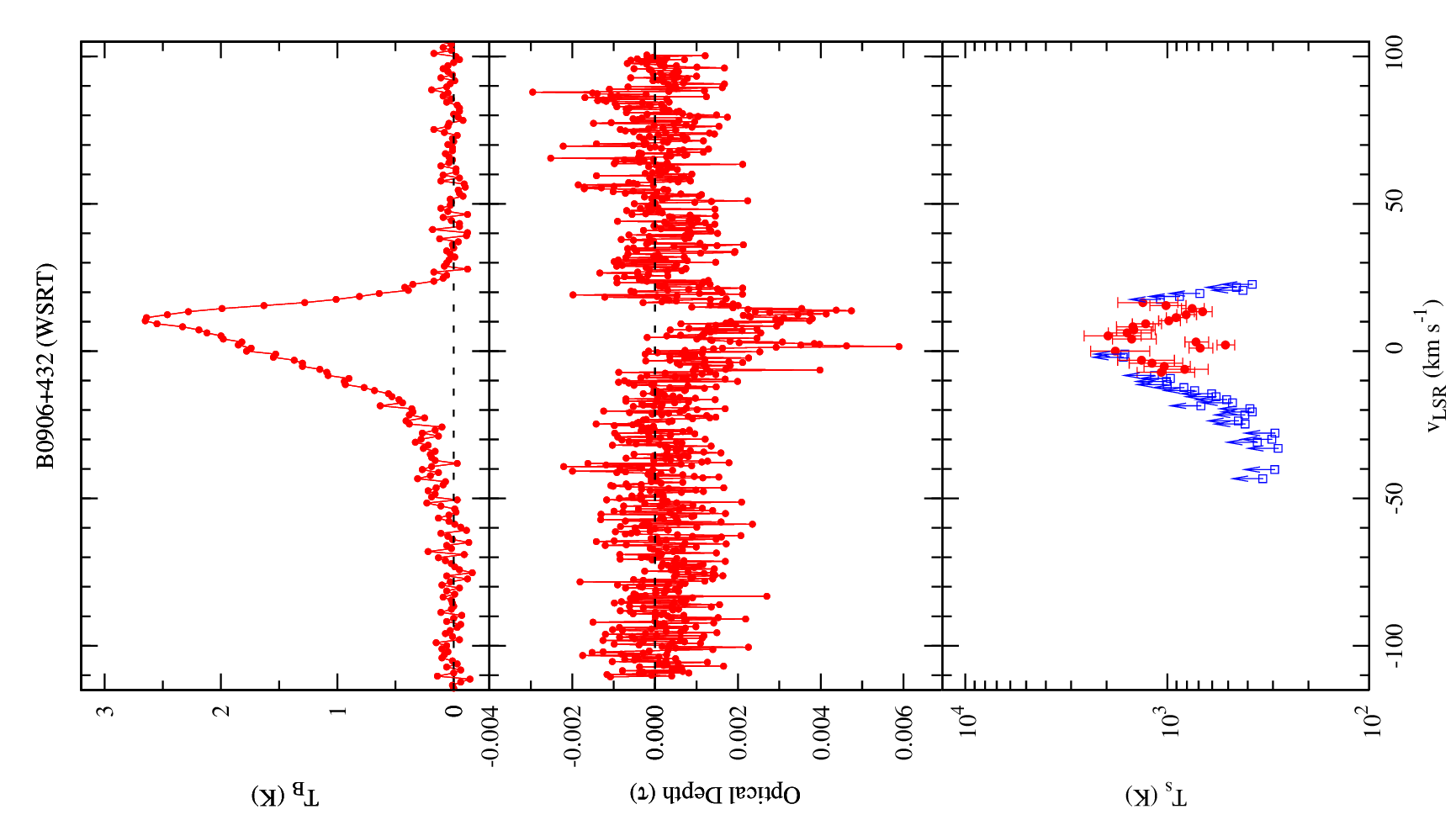}
\includegraphics[height=2.2in,angle=-90]{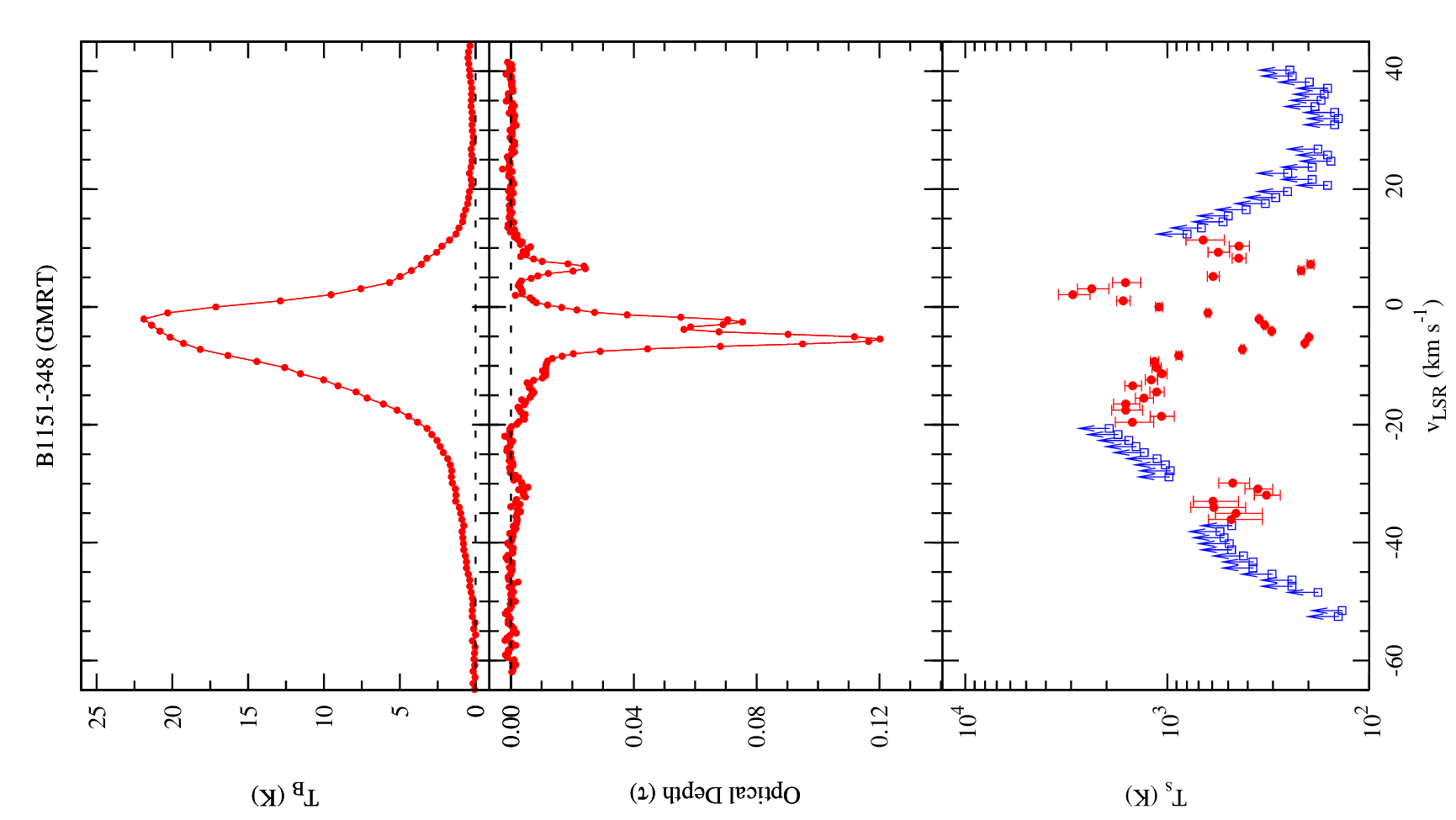}
\includegraphics[height=2.2in,angle=-90]{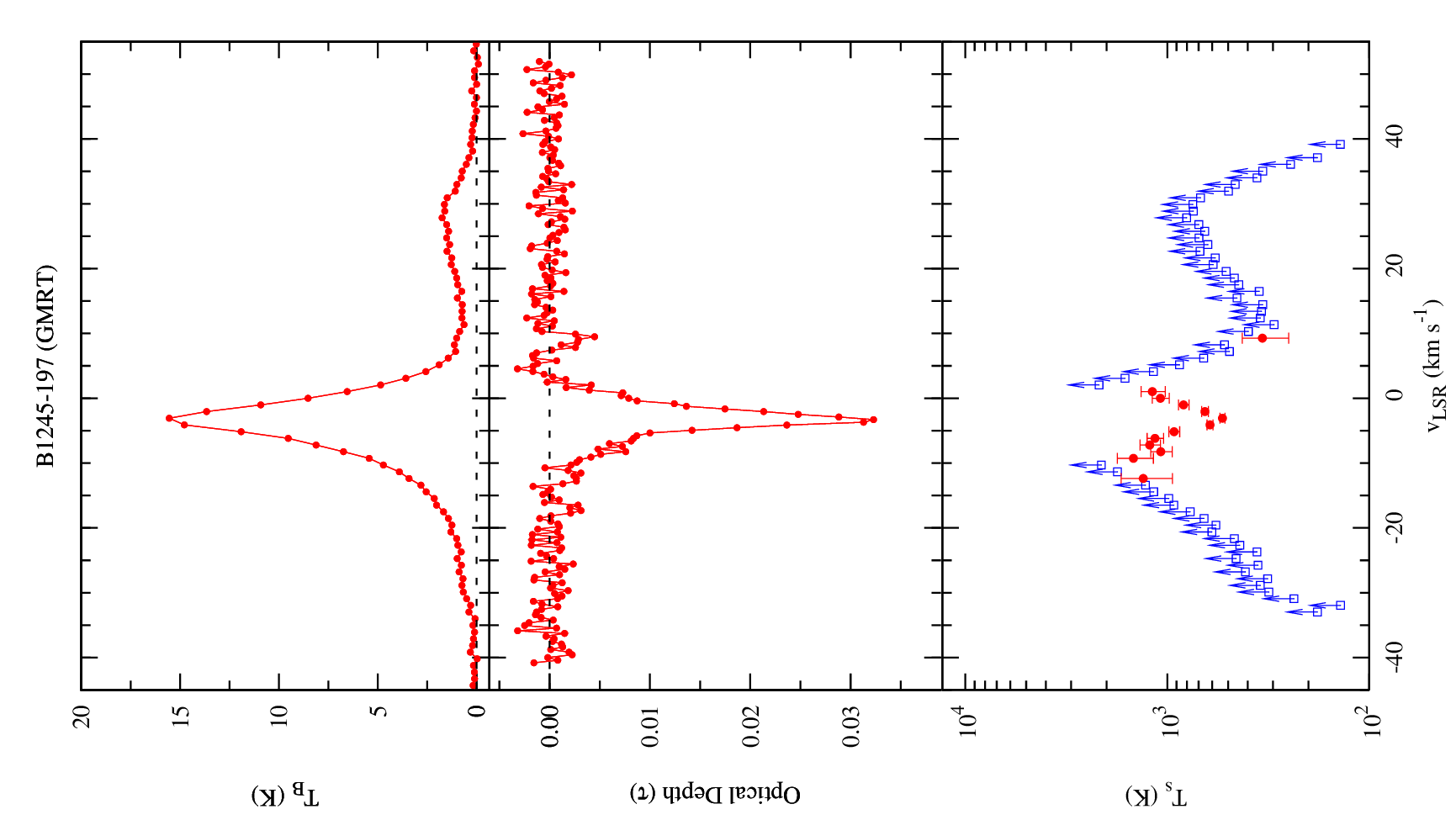}
\includegraphics[height=2.2in,angle=-90]{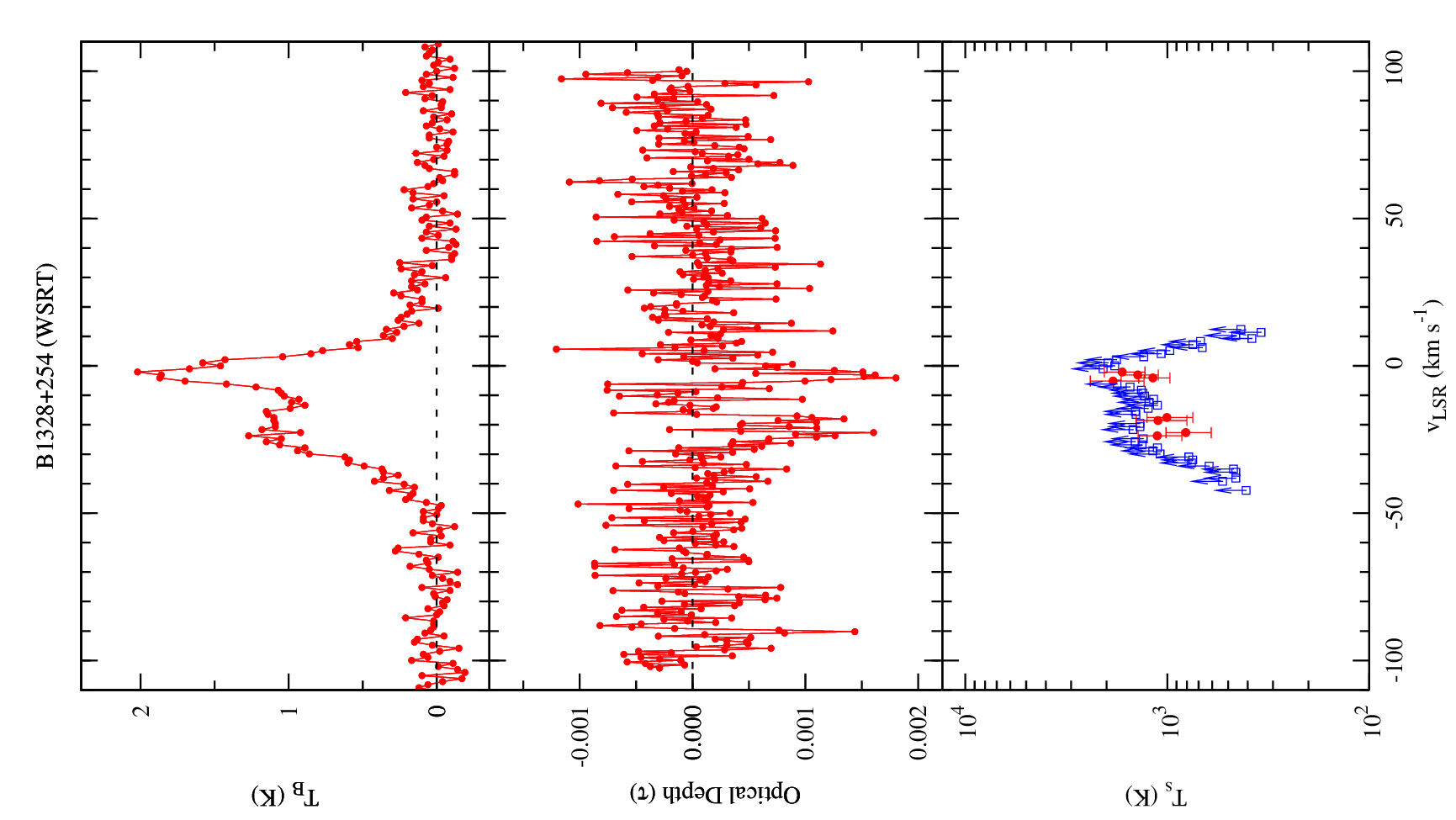}
\includegraphics[height=2.2in,angle=-90]{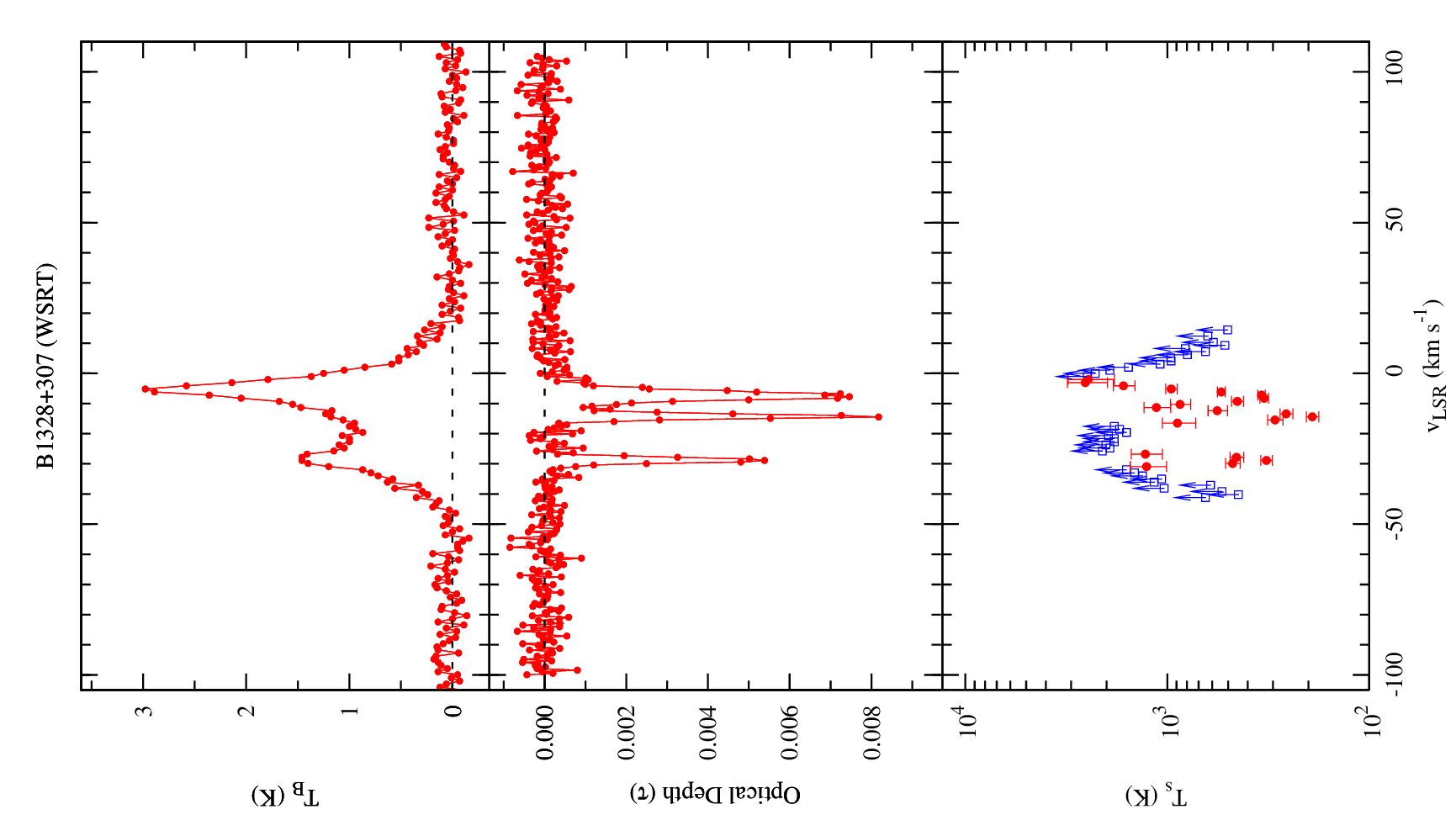}
\includegraphics[height=2.2in,angle=-90]{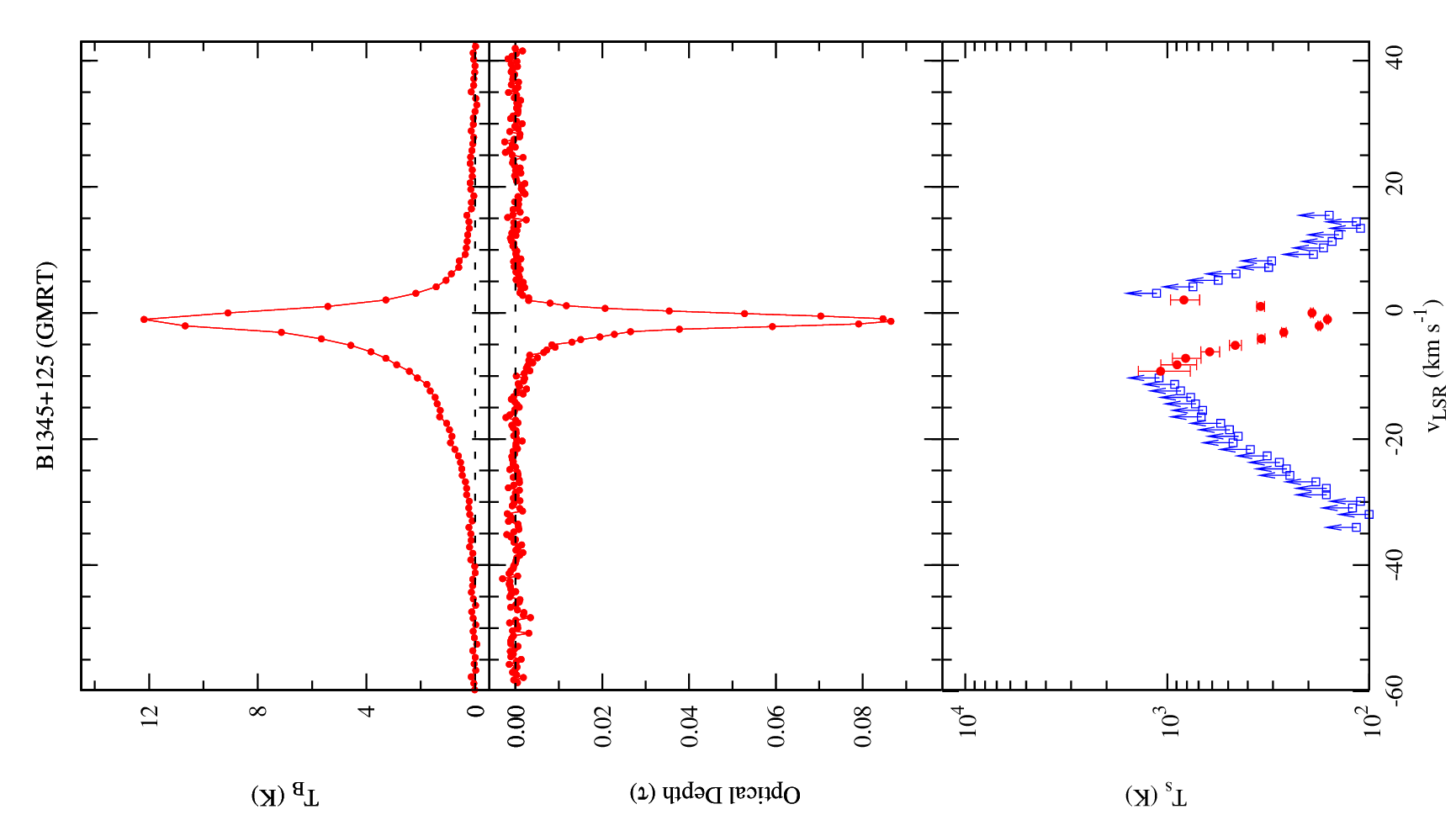}
\caption{\label{fig:spectra-4} ({\it continued}) \hii\ absorption spectra 
obtained using the GMRT/WSRT/ATCA, \hii\ emission spectra from the LAB survey 
and the spin temperature spectra.}
\end{center}
\end{figure*}

\setcounter{figure}{1}
\begin{figure*}
\begin{center}
\includegraphics[height=2.2in,angle=-90]{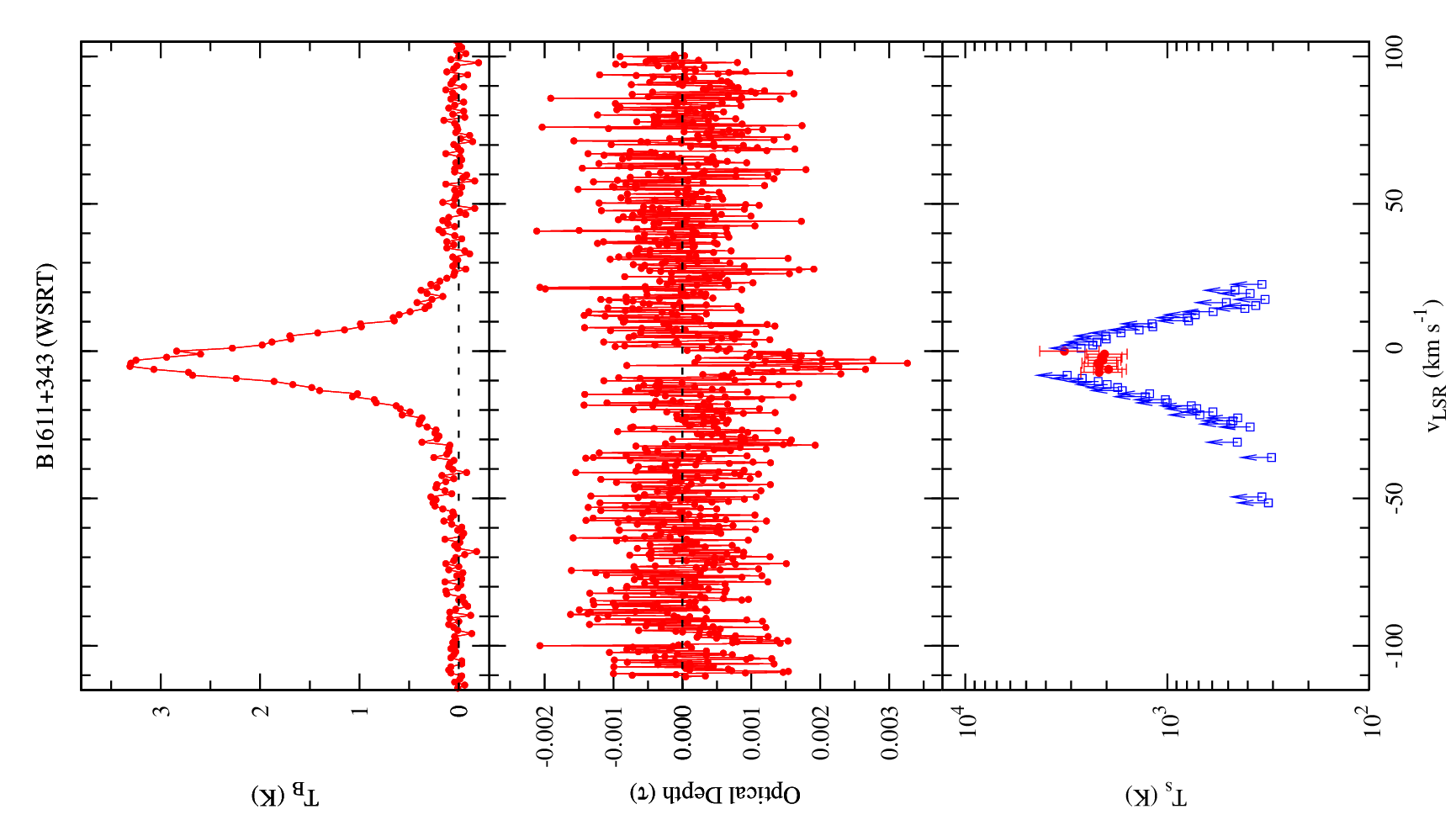}
\includegraphics[height=2.2in,angle=-90]{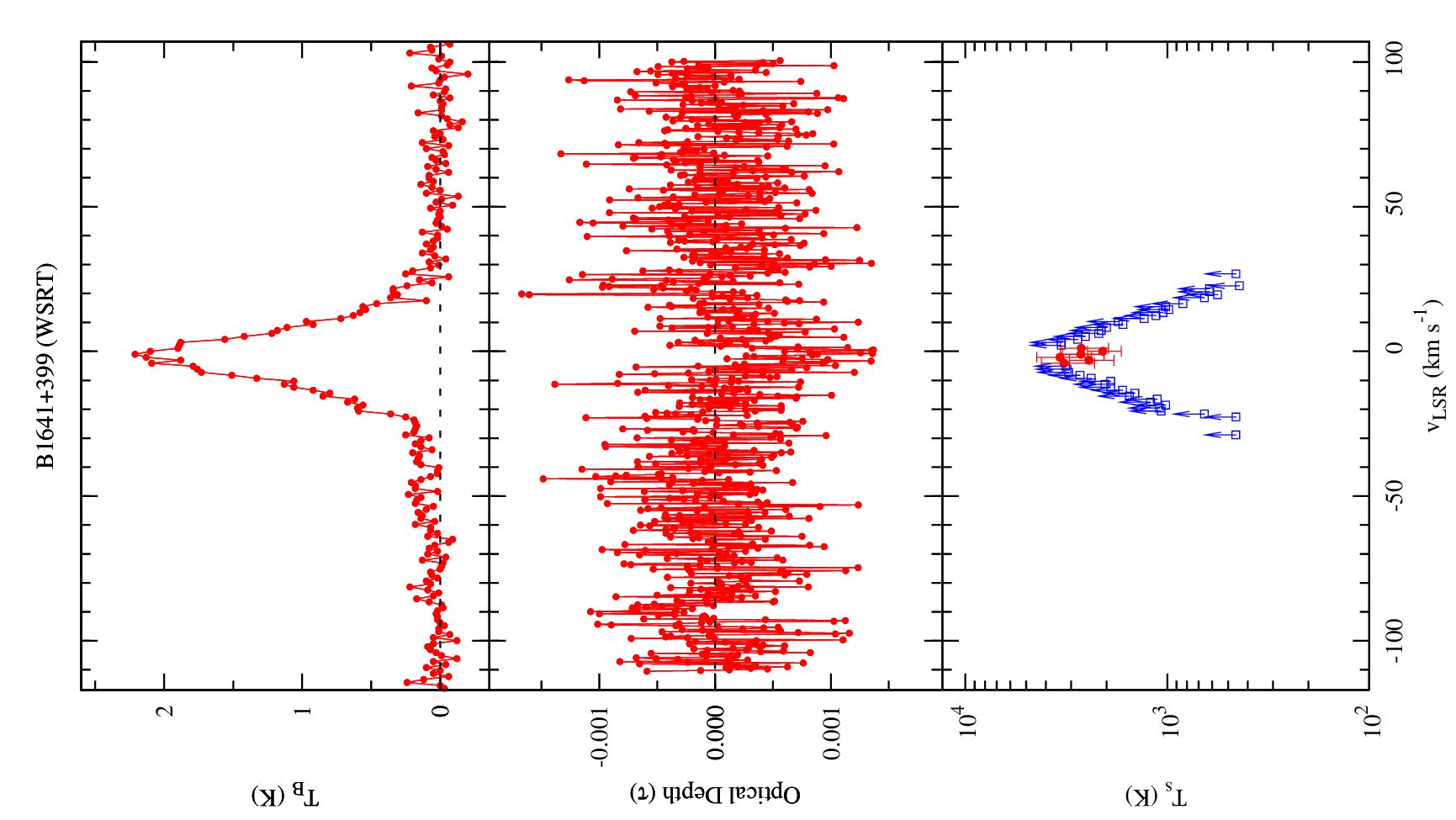}
\includegraphics[height=2.2in,angle=-90]{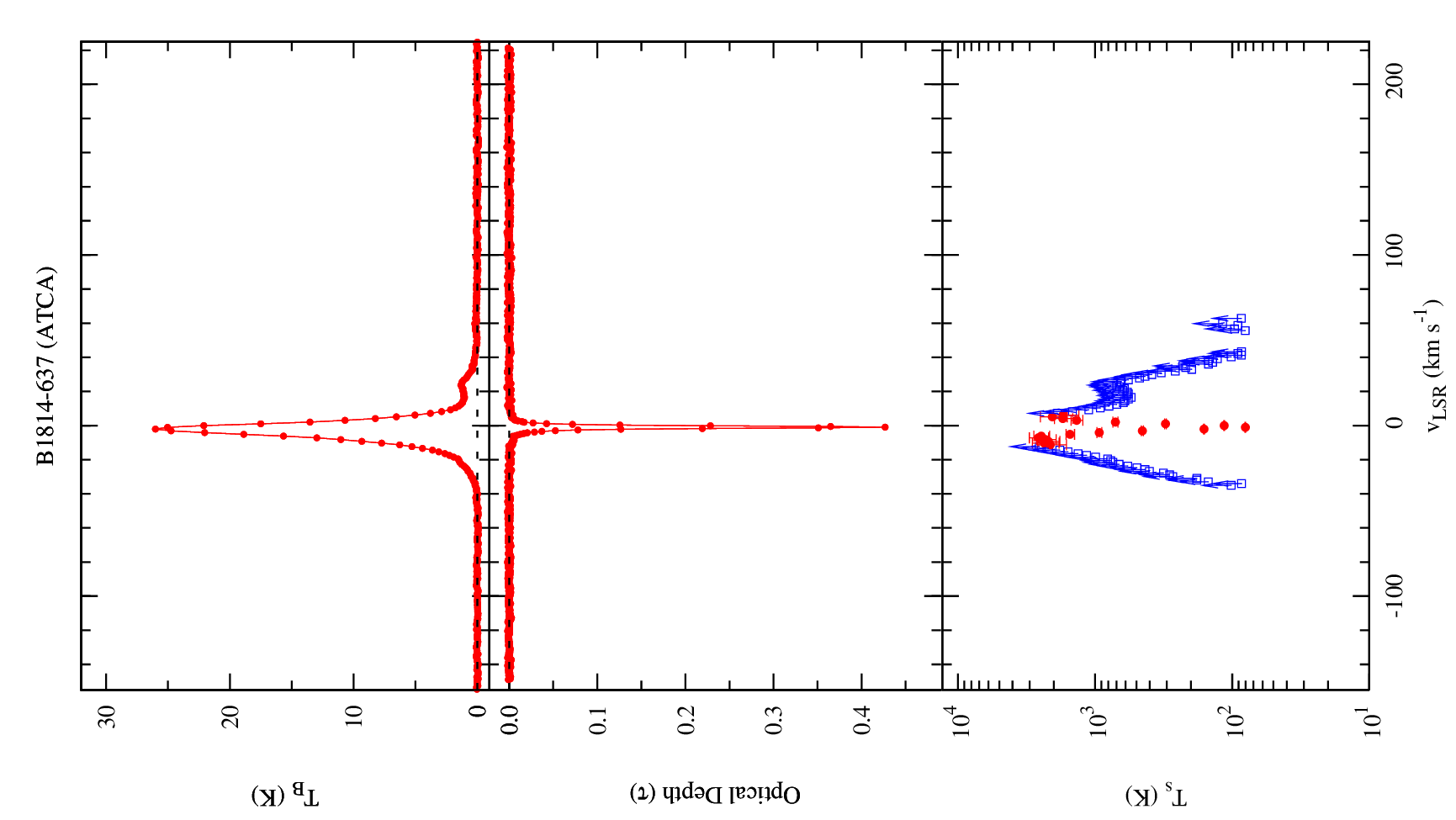}
\includegraphics[height=2.2in,angle=-90]{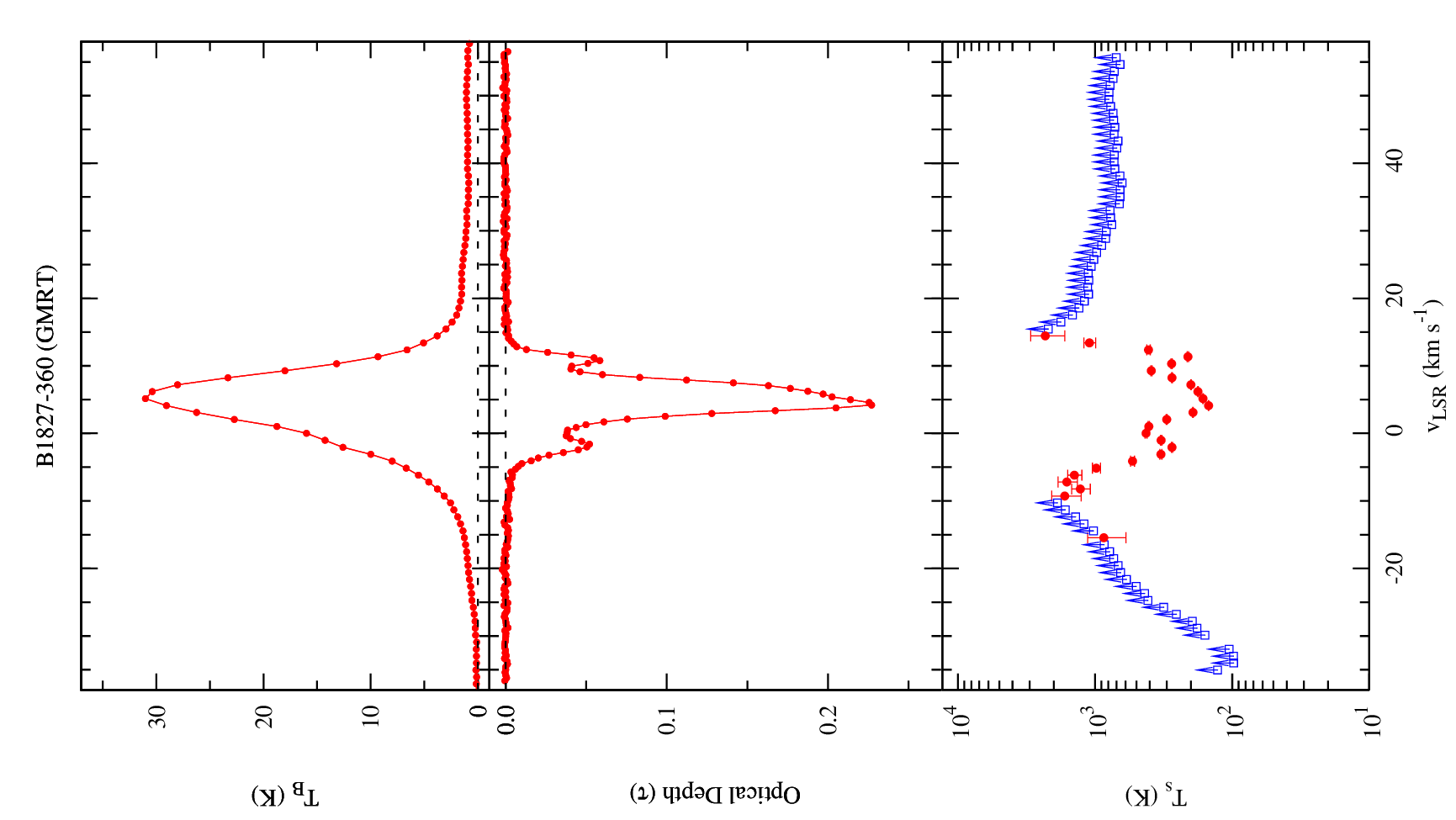}
\includegraphics[height=2.2in,angle=-90]{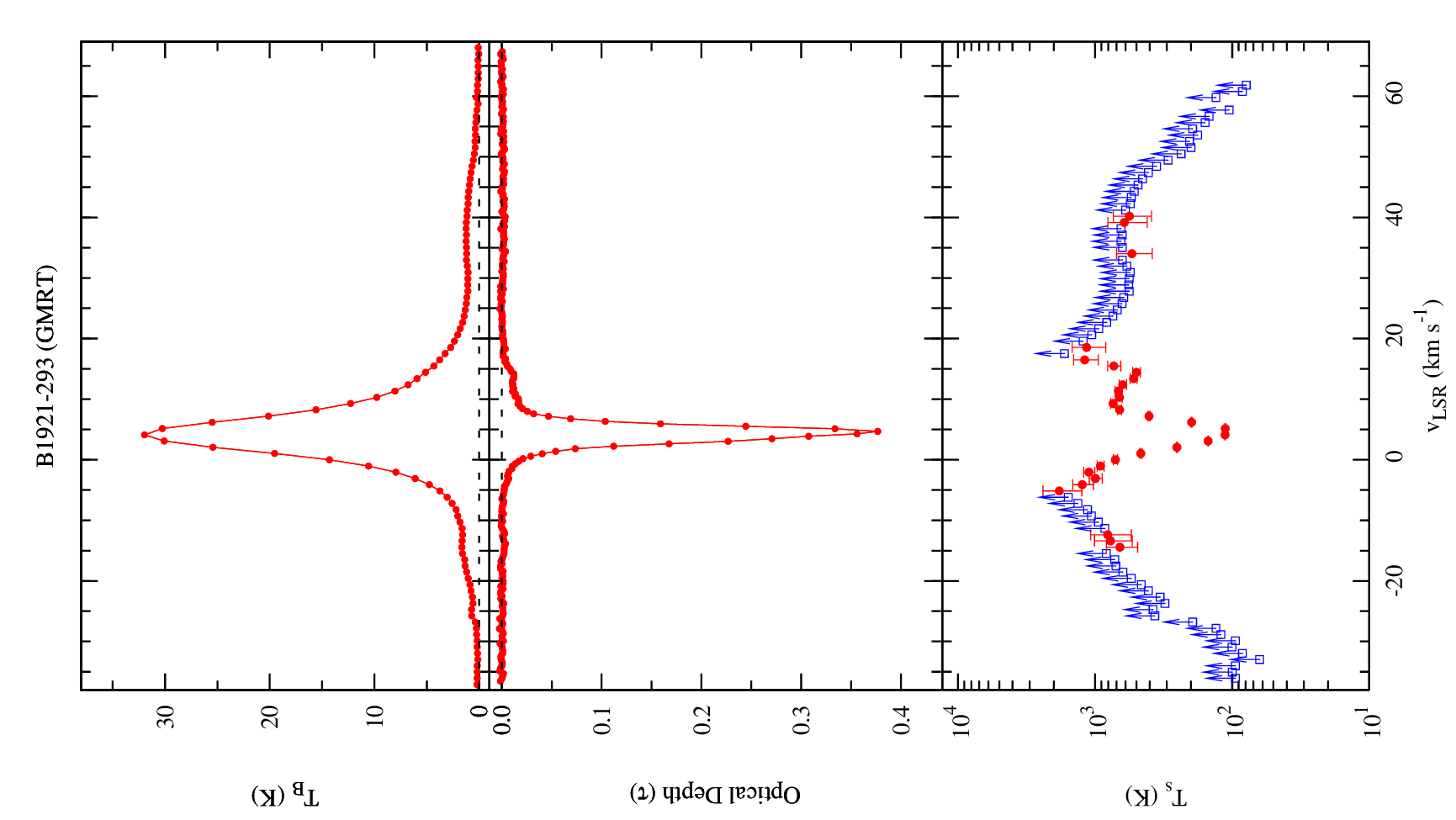}
\includegraphics[height=2.2in,angle=-90]{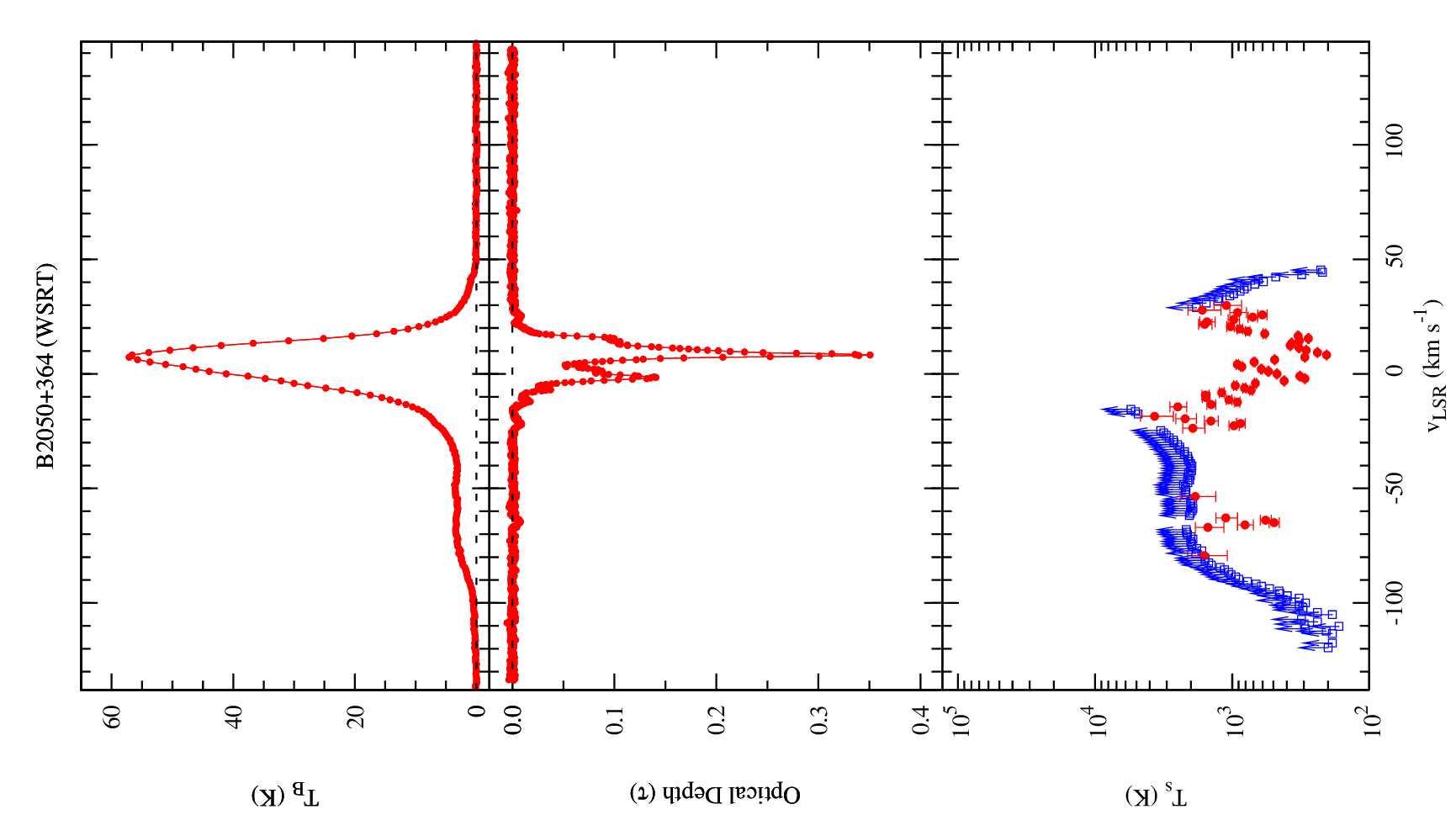}
\caption{\label{fig:spectra-5} ({\it continued}) \hii\ absorption spectra 
obtained using the GMRT/WSRT/ATCA, \hii\ emission spectra from the LAB survey 
and the spin temperature spectra.}
\end{center}
\end{figure*}

\setcounter{figure}{1}
\begin{figure*}
\begin{center}
\includegraphics[height=2.2in,angle=-90]{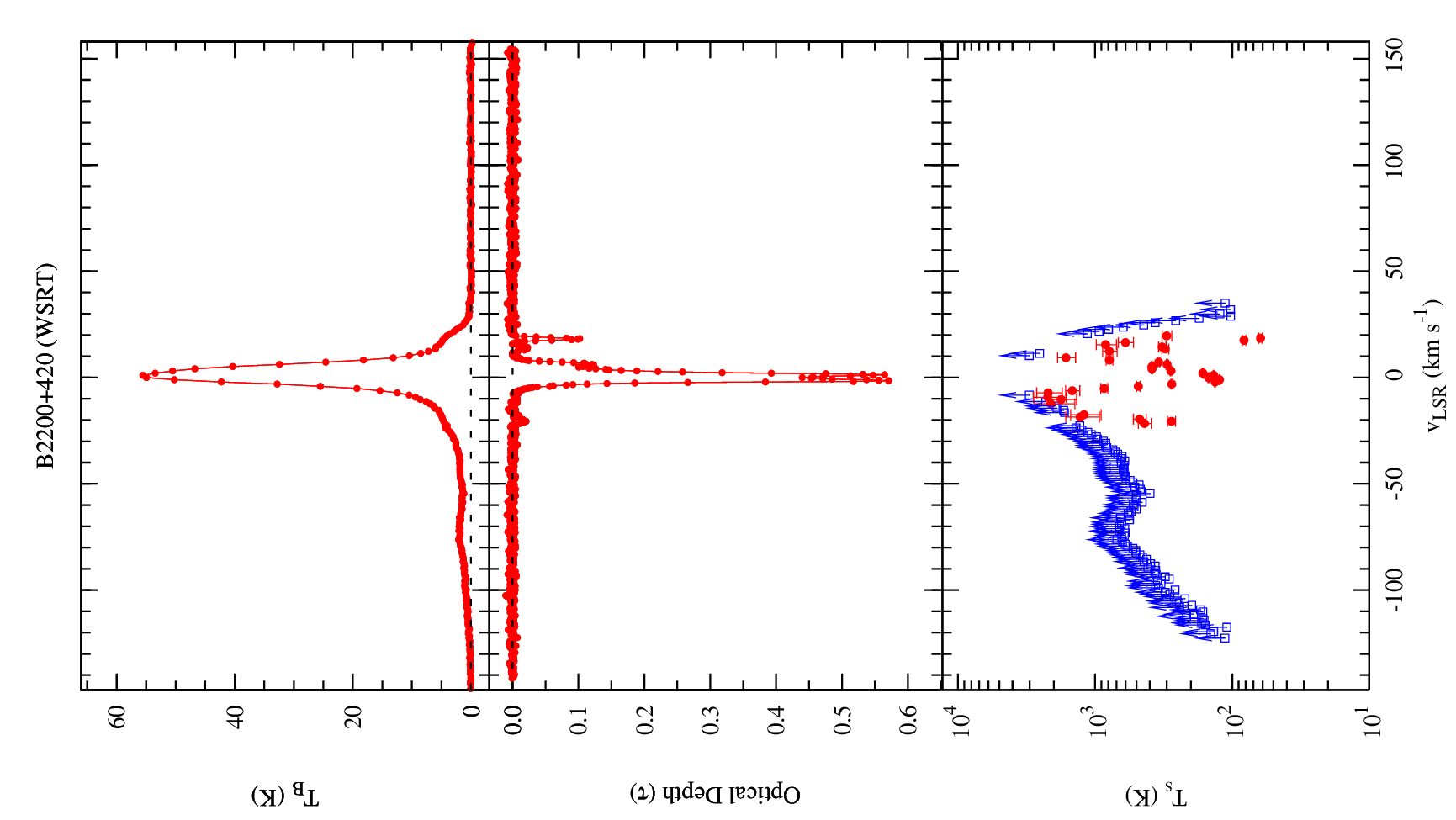}
\includegraphics[height=2.2in,angle=-90]{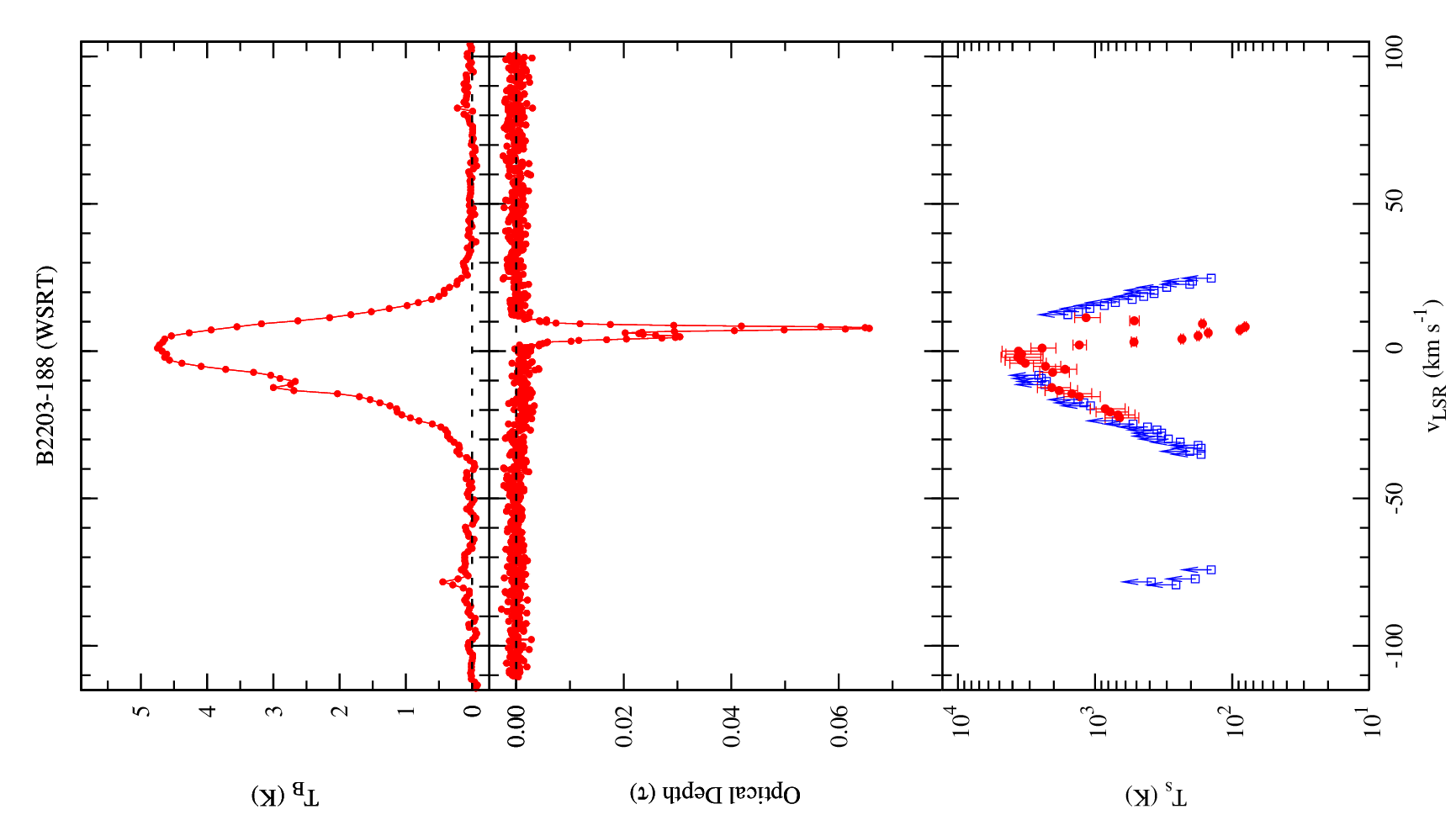}
\includegraphics[height=2.2in,angle=-90]{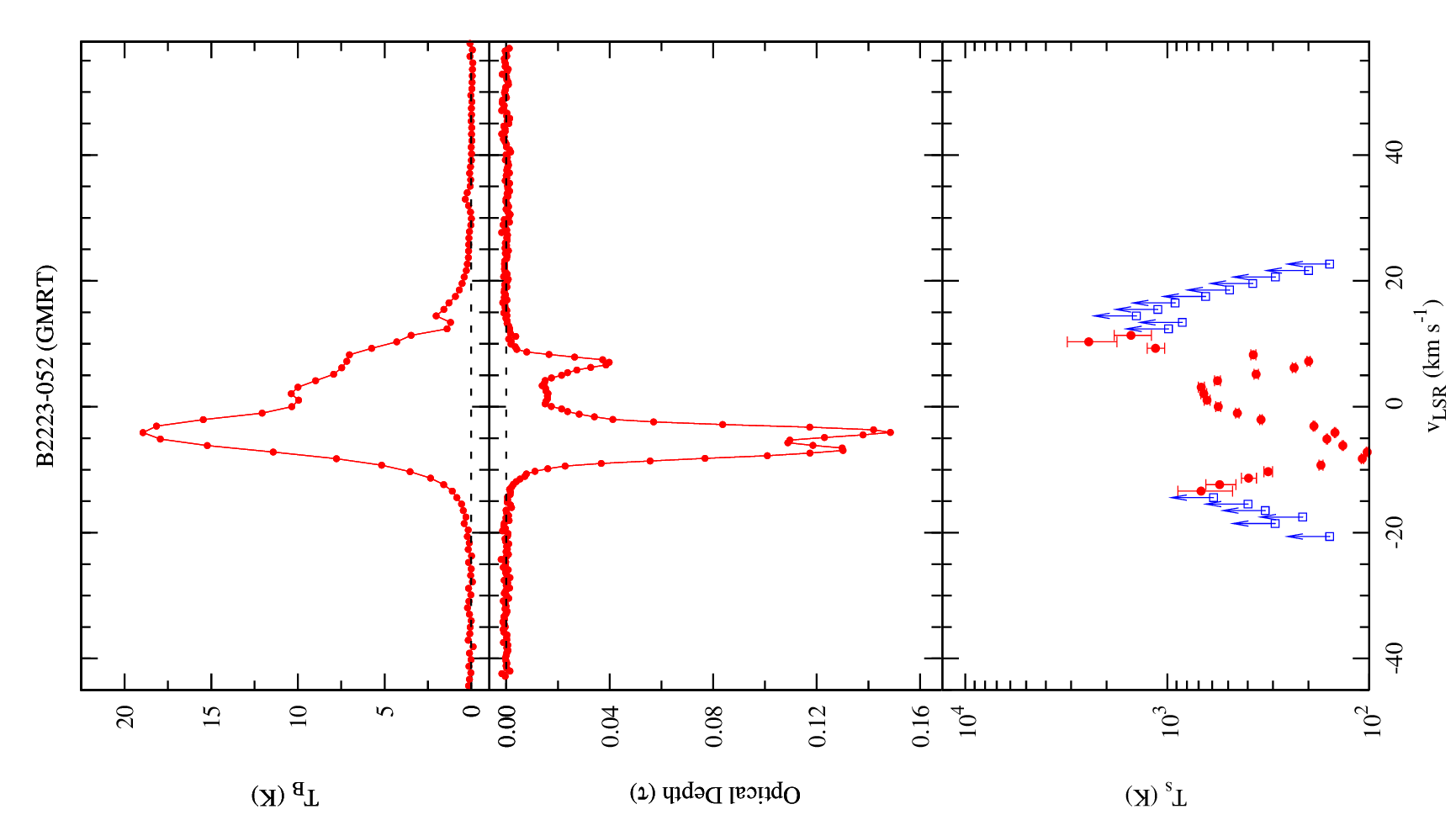}
\includegraphics[height=2.2in,angle=-90]{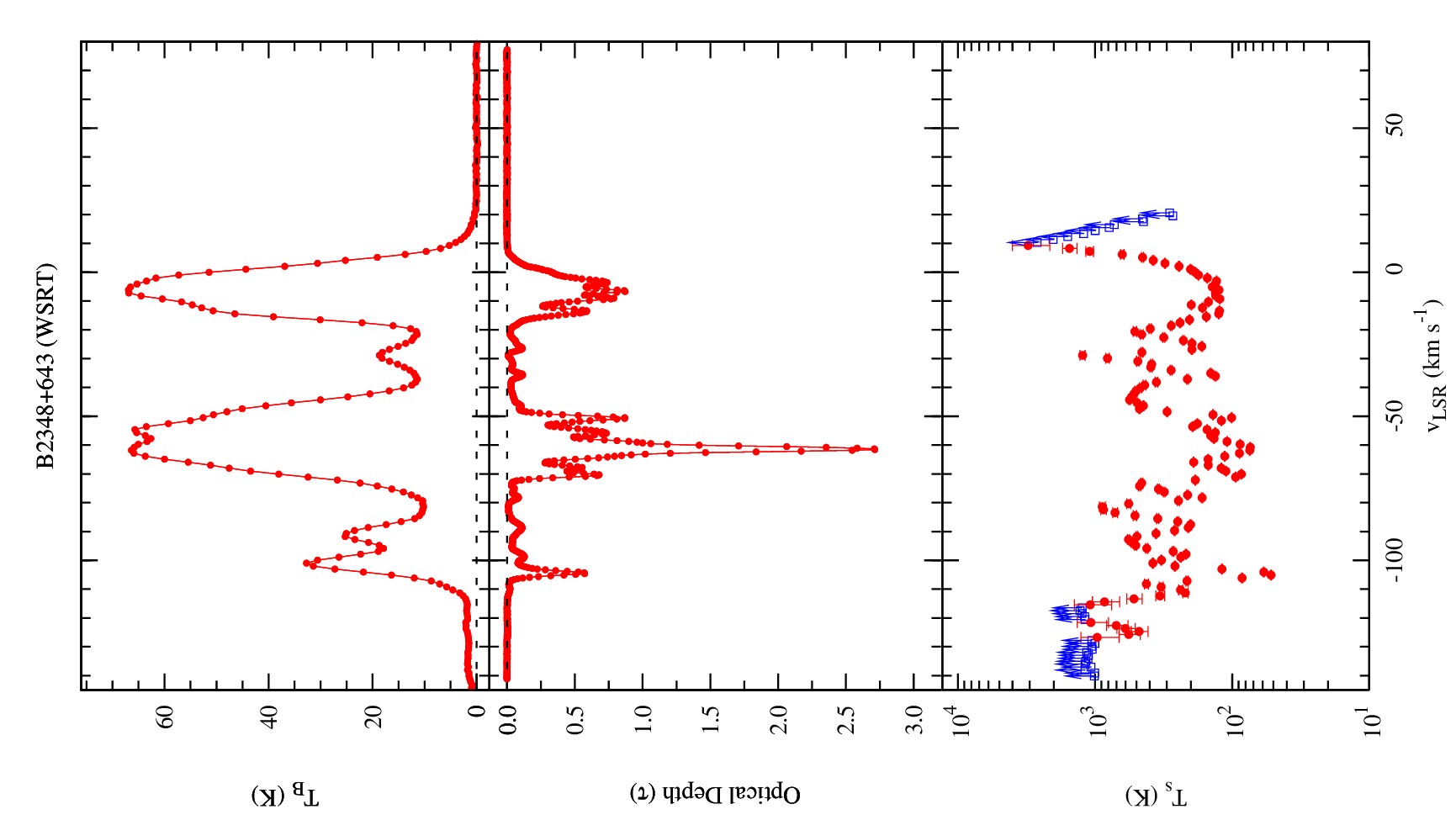}
\caption{\label{fig:spectra-6} ({\it continued}) \hii\ absorption spectra 
obtained using the GMRT/WSRT/ATCA, \hii\ emission spectra from the LAB survey 
and the spin temperature spectra.}
\end{center}
\end{figure*}

The observational results are summarized in Table~\ref{table:results}, whose 
columns contain (1)~the background source name, (2)~its Galactic coordinates, 
(3)~its 1.4~GHz flux density, (4)~the root-mean-square optical depth noise, in 
off-line regions, at resolutions of $0.26$ or $0.52$~\kms\ for the WSRT 
spectra and $0.4$~\kms\ for the GMRT spectra, (5)~the peak \hii\ optical 
depth, (6)~the velocity-integrated \hii\ optical depth $\int\tau{\rm dV}$, in 
\kms, (7)~the total \hi\ column density N(\hi) in the optically-thin limit 
(i.e. a lower limit to the true \hi\ column density along the sightline), 
(8)~the total \hi\ column density N(\hi,ISO) in the isothermal limit 
\citep{chengalur13}, (9)~the column-density-weighted harmonic mean spin 
temperature $\Ts$ obtained using the isothermal estimate of the \hi\ column 
density, and (10-11)~$\Delta V_{90}^{em}$ and $\Delta V_{90}^{abs}$, the 
velocity widths of the profile covering 90\% of the integrated \hii\ emission 
and \hii\ optical depth, respectively. Finally, in the case of B0438$-$436, 
the sole source without detected \hii\ absorption, the columns list the 
$3\sigma$ upper limits on $\tau_{peak}$ and $\int\tau{\rm dV}$, and the 
$3\sigma$ lower limit on $\Ts$ for an assumed line FWHM of 20~\kms, after 
smoothing the spectrum to this resolution.

In passing, the errors on the isothermal estimate of the \hi\ column density 
N(\hi,ISO) are based on the simulations of \citet{chengalur13}, which showed 
that the isothermal limit provides a good estimate of the true \hi\ column 
density with errors of $\approx 30$\% at $\tau \approx 1$ and of a factor of 
$\approx 3$ for large $\tau$ ($68$\% confidence intervals in both cases). We 
have hence computed the \hi\ column density per 1~\kms\ velocity channel using 
three optical depth ranges: (1)~for $\tau < 0.1$, we use the optically-thin 
limit, (2)~for $0.1 \le \tau < 1$, we use the isothermal estimate, with errors 
of $30$\% on the inferred N(\hi,ISO), and (3)~for the few channels with $\tau 
> 1$ (on two sightlines, B0359+508 and B0410+768), we use the isothermal 
estimate, with errors of a factor of $3$ on the inferred N(\hi,ISO). The \hi\ 
column densities per channel are then added together and the errors added in 
quadrature to get the total \hi\ column density along each sightline, quoted 
in column~(8) of Table~\ref{table:results}. As expected, the isothermal 
estimate of the \hi\ column density is somewhat larger than the optically-thin 
estimate only in the case of sightlines with high peak optical depth ($\gtrsim 
1$), whereas the difference is negligible for the majority of the sightlines.

\setcounter{figure}{2}
\begin{figure}
\begin{center}
\includegraphics[height=3.3in]{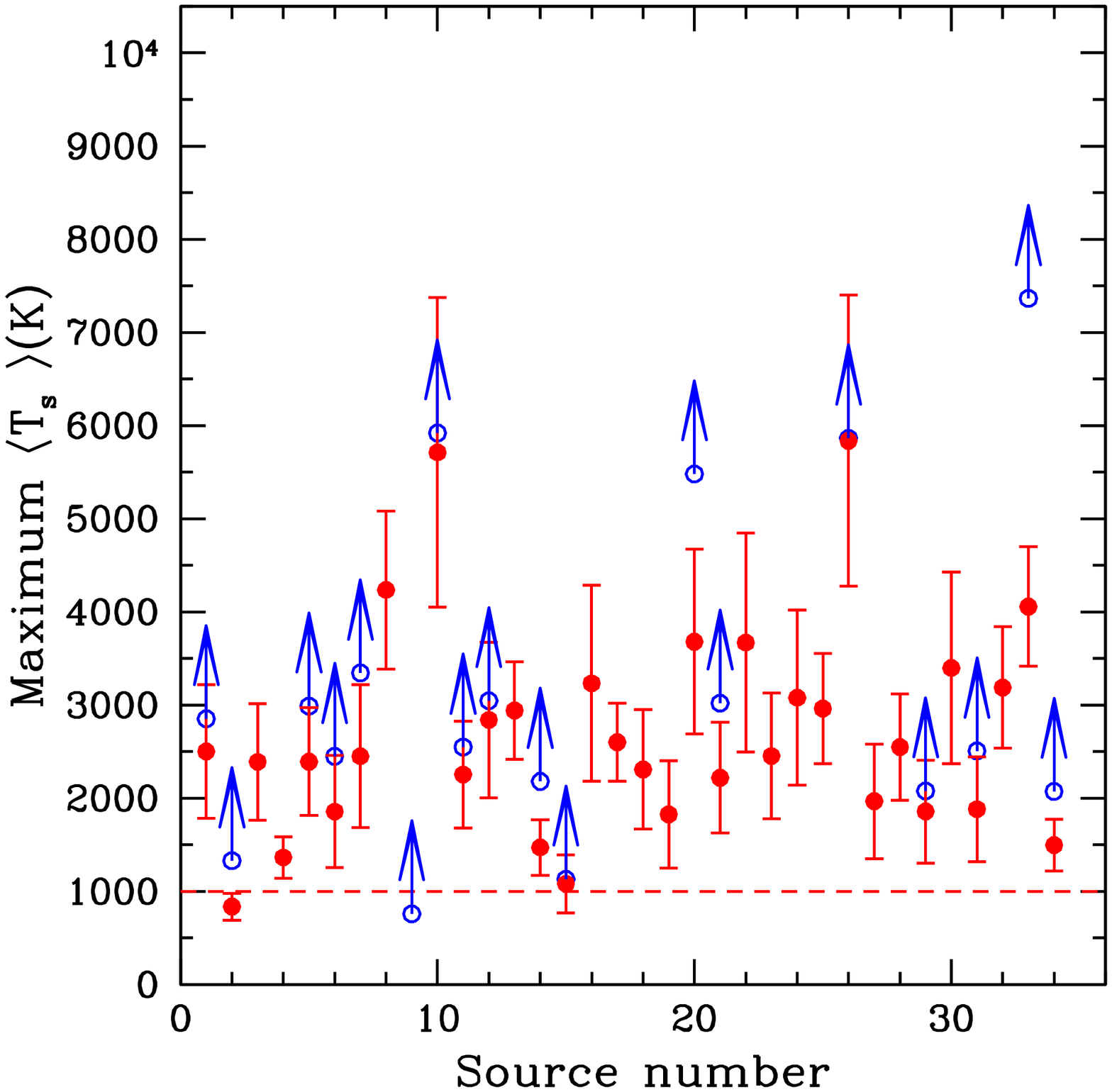}
\caption{\label{fig:tspin} The maximum spin temperature $\Tsmax$ detected at 
$\ge 3\sigma$ significance at $\approx 1$~\kms\ resolution on each sightline 
plotted (in filled circles) versus source number (running from 1 to 34, 
following the order of Fig.~\ref{fig:spectra}). For sightlines with velocity 
channels without detected \hii\ absorption, for which the $3\sigma$ lower 
limit to the spin temperature is higher than $\Tsmax$, we also show the 
highest $3\sigma$ lower limit to $\Ts$ as an open circle and an upward 
pointing arrow. For B0438$-$436, the only source with a non-detection, the 
maximum $3\sigma$ lower limit to the spin temperature is shown. It is clear 
that most sightlines have $\Tsmax \gtrsim 1000$~K.}
\end{center}
\end{figure}

\section{Spin and Kinetic temperatures in the ISM}
\label{sec:tk-ts}

For neutral hydrogen, two different temperatures are used to describe physical 
conditions in the gas: these are the kinetic temperature, \Tk, and the spin 
temperature, ${\rm T_s}$. The spin temperature is the temperature that describes 
the population distribution in the two hyperfine levels of the ground state of 
neutral hydrogen, i.e. it is defined by the Boltzmann equation
\begin{equation}
{n_2 \over n_1} = {g_2 \over g_1} e^{-h\nu_{21}/k{\rm T_s}}
\end{equation}
where $n_1$ and $n_2$ are the populations in, respectively, the lower and upper 
levels of the \hii\ hyperfine transition and $\nu_{21}$ is the frequency of the 
\hii\ line. 
Observationally, ${\rm T_s}$ is estimated by comparing the \hii\ optical depth 
(measured from the absorption spectrum against a compact background continuum 
source) to the total \hi\ column density. The latter is typically obtained by 
measuring the \hii\ brightness temperature along a neighbouring sightline. For 
a single homogeneous cloud, these two measurements allow one to uniquely 
determine the spin temperature and the \hi\ column density. However, a typical 
sightline is likely to contain a mix of gas in different temperature phases. 
For this general situation, it can be shown that the spin temperature derived 
from the above procedure for a line of sight is the column-density-weighted 
harmonic mean of the spin temperatures of different phases along the sightline 
\citep[e.g.][]{kulkarni88}. As such, it does not necessarily provide specific 
spin temperature values for any individual phase. Throughout this paper, we 
use the notation ${\rm T_s}$ for the spin temperature of a single, isothermal 
cloud, and $\Ts$ for the column-density-weighted harmonic mean spin 
temperature along any sightline.

It should be noted that it is the {\it kinetic} temperature that is directly 
predicted by the multi-phase ISM models, while it is typically the {\it spin} 
temperature that is reported in observational studies. For example, in the 
case of the WNM, earlier studies have typically obtained estimates of (or 
lower limits on) the WNM spin temperature \citep[e.g.][]{carilli98b,dwaraka02}. 
It is hence important to clarify that, even for a single \hi\ cloud, the 
kinetic temperature is {\it not} necessarily the same as the spin temperature 
\citep[e.g.][]{field58,kulkarni88,liszt01}. 

For a single \hi\ cloud, ${\rm T_s}$ is determined by a combination of 
collisions, the Ly-$\alpha$ radiation field and the radiation field near the 
\hii\ wavelength \citep{field58}. In the ISM, the \hii\ continuum radiation 
field generally contributes negligibly to the excitation. At high densities 
and in the typical ISM (i.e. far from bright UV sources), the collisional term 
dominates over the Ly-$\alpha$ excitation term and the spin temperature is 
equal to the kinetic temperature; this is the case in the CNM. However, in the 
WNM, the density is not sufficiently high for the \hii\ transition to be 
thermalized by collisions. For example, in the absence of external influences 
(bright radio sources and sources of ultraviolet photons) \citet{liszt01} 
obtained  ${\rm T_s} \sim 1000 - 5000$\,K for $\Tk \sim 5000 - 10000$\,K, for 
typical ISM pressures. The importance of excitation by the Lyman-$\alpha$ in 
the WNM is somewhat uncertain. If the Lyman-$\alpha$ excitation is 
sufficiently intense and resonant scattering of Lyman-$\alpha$ radiation 
causes the Lyman-$\alpha$ colour temperature to approach the kinetic 
temperature \citep{wouthuysen52,field59}, then the spin temperature would be 
driven to the kinetic temperature.  On the other hand it is not clear that 
(i)~in the three-phase McKee-Ostriker model \citep{mckee77}, photons from the 
Galactic UV background penetrate sufficiently deep in the WNM and (ii)~even if 
the WNM is threaded by sufficient Lyman-$\alpha$ radiation, the colour 
temperature of the Lyman-$\alpha$ radiation would be equal to the local 
kinetic temperature. \citet{liszt01} suggests that it is unlikely that the 
sufficient amount of Lyman-$\alpha$ radiation penetrates the WNM to thermalize 
it, and that in general one would expect that ${\rm T_s}$ in the WNM would be 
lower than $\Tk$. It thus appears reasonable to assume that ${\rm T_s} \approx 
\Tk$ in the CNM and that ${\rm T_s} < \Tk$ in the WNM. We emphasize that these
statements are valid for a single homogeneous \hi\ cloud.

Next, a typical sightline contains a mix of gas at different temperatures.
The fact that the inferred spin temperature for such a sightline is a 
column-density-weighted harmonic mean of the spin temperatures of the different
phases implies that it is biased towards the cold phase. Indeed, a sightline with 
gas equally divided between cold gas at ${\rm T_s} = 100$~K and warm gas at 
${\rm T_s} = 8000$~K 
would yield an inferred harmonic-mean spin temperature of $\Ts \approx 200$~K. 
Conversely, even if 90\% of the gas has ${\rm T_s} = 8000$~K, while the rest 
has ${\rm T_s} = 100$~K, the inferred harmonic mean spin temperature would be 
$\Ts \approx 900$~K. Note that 8000~K is around the upper limit to the WNM 
kinetic temperature and ${\rm T_s}$ is expected to be less than $\Tk$ in the 
WNM; hence, WNM spin temperatures are expected to be significantly lower than 
8000~K (e.g. $\approx 1000-4000$~K in the models of \citealt{liszt01}). Thus, 
a harmonic-mean spin temperature of $\approx 1000$~K can only arise if almost 
all the gas is in the WNM. In summary, a low inferred harmonic-mean spin 
temperature ($\lesssim 300$~K) implies that $\gtrsim 50$~\% of the gas along 
the sightline is in the CNM, while a high inferred harmonic-mean spin 
temperature ($\gtrsim 1000$~K) implies that the gas is almost entirely in the 
WNM. 

The line-of-sight harmonic mean spin temperature values for the sources of 
our sample are listed in Table~\ref{table:results}. The lowest panels of 
Fig.~\ref{fig:spectra} also show the harmonic-mean spin temperature 
``spectra'' for the 34 sightlines of the full interferometric sample (i.e. 
including the GMRT, WSRT and ATCA data). These were obtained by smoothing the 
\hii\ absorption spectra to the LAB velocity resolution, and then obtaining 
the spin temperature from the relation $\Ts = \Tb/(1 - e^{-\tau})$. The points 
with errors indicate $\Ts$ measurements with $\ge 3 \sigma$ significance at a 
velocity resolution of $\approx 1$~\kms, while the arrows indicate $3\sigma$ 
lower limits on the spin temperature. It is clear that most sightlines have 
$\Ts$ measurements that are $\gtrsim 1000$~K. This is shown explicitly in 
Fig.~\ref{fig:tspin} which plots the maximum spin temperature $\Tsmax$ 
detected at $3\sigma$ significance and the highest $3\sigma$ lower limit to 
the spin temperature on each sightline against source number (from 
Fig.~\ref{fig:spectra}). The single sightline with undetected \hii\ absorption, towards B0438$-$436, is shown as a $3\sigma$ lower limit to $\Ts$. Most 
sightlines have $\Tsmax \gtrsim 1000$~K, indicating (see above) that the gas 
is almost entirely in the WNM; indeed, roughly 75\% of the sources have 
$\Tsmax \ge 2000$~K. 

\setcounter{figure}{3}
\begin{figure*}
\begin{center}
\includegraphics[height=7in,angle=-90]{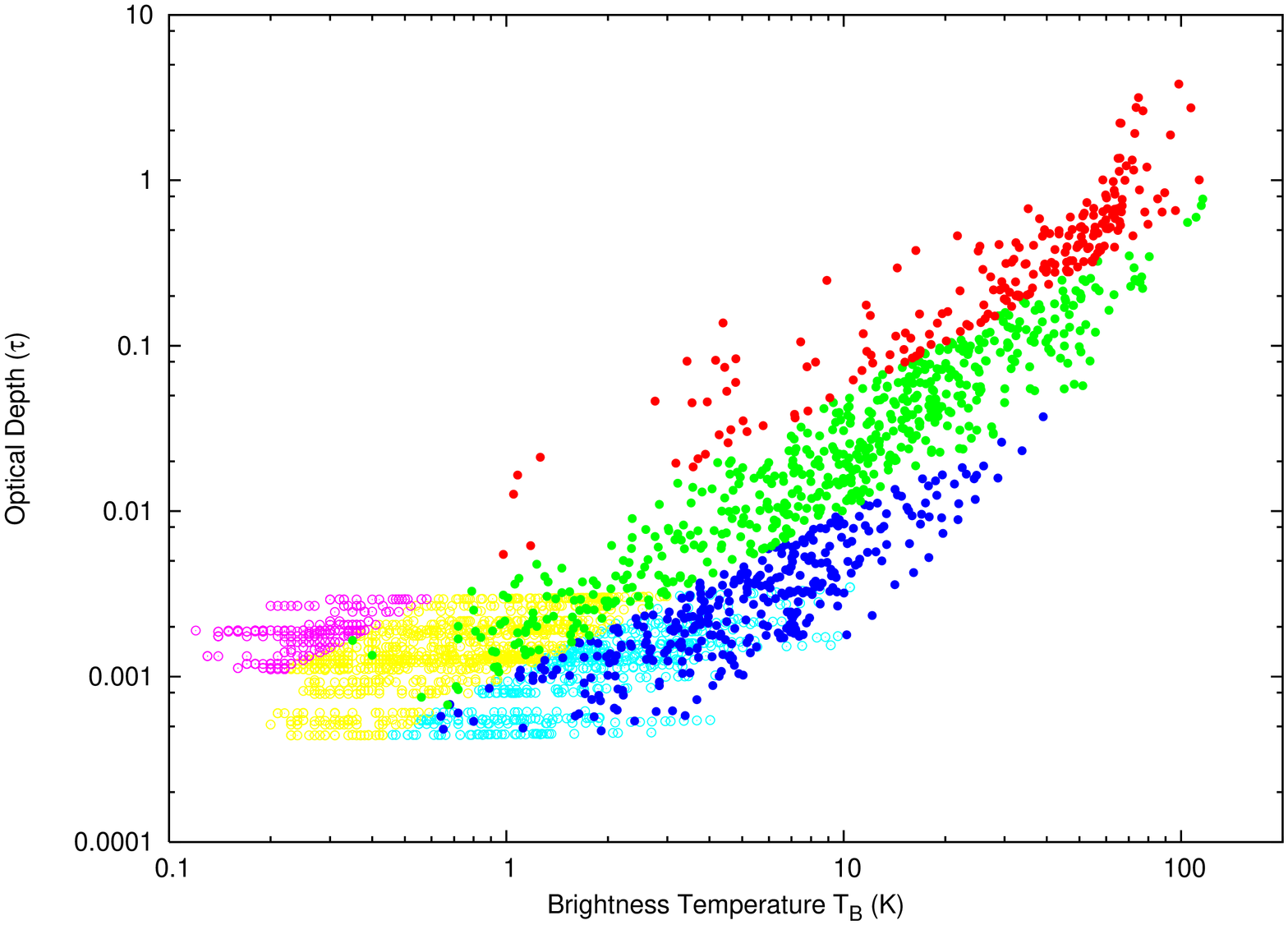}
\caption{\label{fig:tau-tb-ts} An overview of the observational data: the 
figure shows the \hii\ optical depth measured at a resolution of 1~\kms\ 
plotted versus the \hii\ brightness temperature for all 1~\kms\ velocity 
channels with \hii\ emission detected at $\ge 3\sigma$ significance. 
Detections are shown as filled circles, and non-detections as open circles. }
\end{center}
\end{figure*}

Following \citet{lazareff75}, a number of \hii\ absorption/emission studies in 
the literature have attempted to make progress by comparing the optical depth 
$\tau$ with the spin temperature $\Ts$, to derive a relation between the two 
quantities. However, as emphasized by \citet{braun92}, the problem actually 
has three dimensions, the \hii\ optical depth $\tau$, the \hii\ brightness 
temperature $\Tb$ and the spin temperature $\Ts$. Fig.~\ref{fig:tau-tb-ts} 
provides an overview of the data in all three dimensions by plotting the \hii\ 
optical depth versus the \hii\ brightness temperature (i.e. the two 
observables), for all the 1~\kms\ velocity channels with detected \hii\ 
emission at $\ge 3\sigma$ significance. Different spin temperature ranges are 
shown in the figure in different colours: the range $\Ts < 200$~K is shown in 
red (detections) and magenta (non-detections, with $3\sigma$ lower limits 
within this range), $200$~K~$ < \Ts < 1000$~K in green (detections) and yellow 
(non-detections), and $1000$~K~$ < \Ts < 10000$~K in blue (detections) and 
cyan (non-detections). The first three ranges correspond, broadly (and for a 
single temperature cloud), to the cold neutral medium, the unstable neutral 
medium and the warm neutral medium, respectively, using the model of 
\citet{liszt01} to relate kinetic temperature to spin temperature. Note that 
spin temperatures larger than $\approx 5000$~K are not expected to be 
typically present even in the WNM \citep[e.g.][]{liszt01}; only a small 
fraction ($\approx 0.4$\%)  of velocity channels have $\Ts \gtrsim 5000$~K. 

It is clear from the figure that velocity channels with $\Tb > 10$~K are 
always detected in absorption when the optical depth detection sensitivity is 
$\gtrsim 0.001$. This is somewhat larger than the limiting brightness 
temperature for detectable \hii\ absorption quoted by \citet{braun92} 
($\approx 5$~K for an optical depth detection sensitivity of $\approx 0.01$). 
It is also apparent that there are few low or intermediate optical depth 
measurements (i.e. high $\Ts$ values) at high brightness temperatures $\Tb >> 
50$~K, although such optical depths would have been easily detected at our 
sensitivity. Clearly, high brightness temperatures are associated with the 
cold neutral medium. More curious is the fact that only a small fraction of 
the gas at low and intermediate brightness temperatures ($\Tb < 10$~K) has low 
spin temperatures, $\Ts < 200$~K, although these would, of course, be easiest 
to detect. 

\setcounter{figure}{4}
\begin{figure*}
\begin{center}
\includegraphics[height=3.4in,angle=-90]{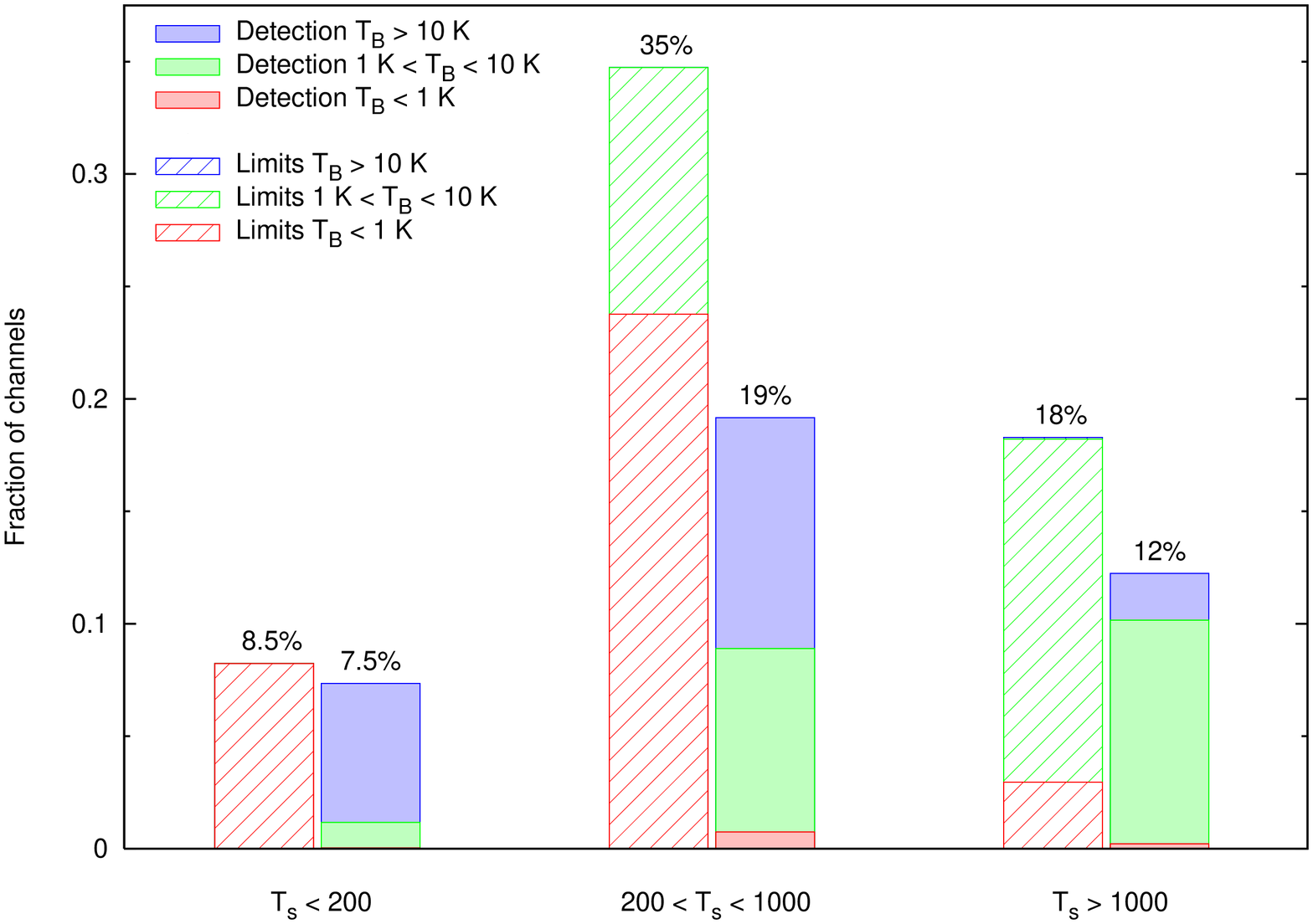}
\includegraphics[height=3.4in,angle=-90]{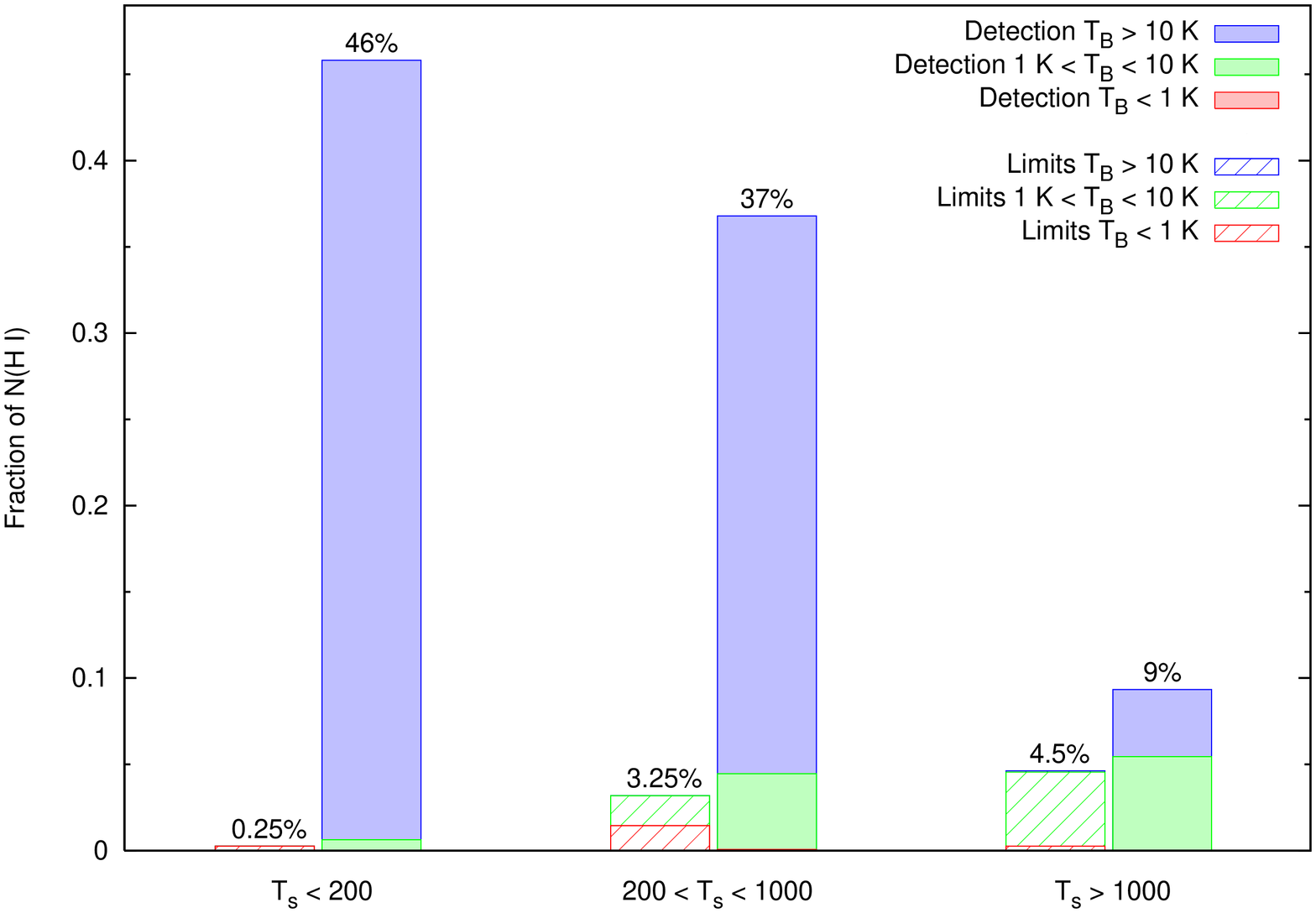}
\caption{\label{fig:histogram} [A]~Left panel: A histogram of the fraction of 
channels from Fig.~\ref{fig:tau-tb-ts} showing \hii\ absorption detections and 
non-detections, for three spin harmonic mean temperature ranges, $\Ts < 
200$~K, $200$~K~$< \Ts < 1000$~K and $\Ts > 1000$~K. [B]~Right panel: A 
similar histogram showing the fraction of \hi\ column density with detections 
and non-detections of \hii\ absorption, again for for three harmonic mean spin 
temperature ranges, $\Ts < 200$~K, $200$~K~$< \Ts < 1000$~K and $\Ts > 
1000$~K. In both panels, detections are shown as solid boxes and 
non-detections as hashed boxes. The red, green and blue regions indicate three 
ranges of brightness temperature, $\Tb < 1$~K, $1$~K~$< \Tb < 10$~K and $\Tb 
> 10$~K, respectively.}
\end{center}
\end{figure*}

Fig.~\ref{fig:histogram}[A] shows a histogram of the results for the above 
spin temperature ranges, comparing the fraction of channels with detections 
and non-detections of \hii\ absorption at different $\Ts$ ranges. For each 
$\Ts$ range, the detection fraction is shown in the solid box, while the 
non-detections with $3\sigma$ lower limits within this range are indicated by 
the hashed box. Within each box, the red, green and blue colours indicate 
brightness temperatures $\Tb < 1$~K, $1$~K~$ < \Tb < 10$~K and $\Tb > 10$~K, 
respectively. Note that the absence of hashed blue regions among the 
non-detections is because \hii\ absorption is always detected for $\Tb > 
10$~K. However, it is curious that there are no detections with $\Ts < 200$~K 
and $\Tb < 1$~K; such gas would have $\tau \approx 0.01$ which would have been 
easily detectable in the survey. The total \hi\ column density of such 
``clouds'' would be only $few \times 10^{18}$~\cm; it is likely that such low 
column density clouds cannot self-shield against the penetration of external 
ultraviolet radiation and hence contain predominantly warm gas 
\citep{kanekar11b}. Roughly one-third of the velocity channels (including 
detections and non-detections) have spin temperatures $> 1000$~K, i.e. clearly 
lie in the WNM range of spin temperatures. 

Fig.~\ref{fig:histogram}[B] shows a similar histogram, but for the fraction of 
\hi\ column density (obtained via the isothermal estimate for each $1$~\kms\ 
channel; \citealt{chengalur13}) with detections and non-detections of \hii\ 
absorption. It is clear that the bulk of the neutral hydrogen ($\approx 92$\%) 
along the 34 sightlines is detected in \hii\ absorption. Slightly less than 
half ($\approx 46$\%) of the \hi\ has $\Ts < 200$~K, $\approx 37$\% has 
$200$~K~$< \Ts < 1000$~K, and only $\approx 9$\% of the \hi\ has $\Ts$ in the 
WNM range of spin temperatures, $\Ts > 1000$~K. We emphasize that this does 
{\it not} mean that half the neutral gas is in the cold phase, $\approx 40$\% 
in the unstable phase and $\approx 10$\% in the warm phase, as the spin 
temperature for each velocity channel is the harmonic mean of the spin 
temperatures of the different phases that contribute to the \hii\ absorption 
and emission.

\section{Summary}
\label{sec:conclude}

We have presented deep interferometric (GMRT, WSRT) absorption spectra of 
Galactic \hii\ absorption towards 32 sources. The sensitivity of the 
observations was sufficient to detect gas in the WNM with column density $\sim 
10^{20}$cm$^{-2}$ and velocity widths comparable to the thermal width. Spectra 
of the same sources taken obtained at the GMRT and WSRT show excellent 
agreement -- this indicates that spectral baseline problems and contamination 
from \hi\ emission are negligible. For each source, we combine our absorption 
spectrum with the \hii\ emission spectrum along a neighbouring sightline from 
the Leiden-Argentine-Bonn survey to derive a spin temperature spectrum. In all 
cases, the maximum spin temperature detected (at $\ge 3\sigma$ significance 
per 1~\kms\ channel) is $\gtrsim 1000$~K, indicating absorption by the warm 
neutral medium. We conclude that the sensitivity of the absorption spectra is 
sufficient to detect absorption by the WNM. Later papers will discuss the 
decomposition of the absorption and the emission spectra into components to 
probe physical conditions in the diffuse interstellar medium.

\section*{Acknowledgements}

This research has made use of the NASA's Astrophysics Data System. We thank 
the staff of the GMRT and the WSRT who have made these observations possible. 
The GMRT is run by the National Centre for Radio Astrophysics of the Tata 
Institute of Fundamental Research. The WSRT is operated by ASTRON (the 
Netherlands Institute for Radio Astronomy), with support from the Netherlands 
Foundation for Scientific Research (NWO). Some of the data used in this paper 
were obtained from the Leiden/Argentine/Bonn Galactic \hi\ Survey. We thank 
James Urquhart for his comments on an earlier version of the paper. NR 
acknowledges the Jansky Fellowship Program of NRAO/NSF/AUI and support from 
the Alexander von Humboldt Foundation. NR also acknowledges support from NCRA 
during his stay there, when a major fraction of this work was done. NK 
acknowledges support from the Department of Science and Technology, through a 
Ramanujan Fellowship.

\bibliographystyle{mn2e}
%\bibliography{ms}

\bsp

\label{lastpage}

\end{document}